\documentclass[11pt]{article}
\pdfoutput=1
\usepackage{amsmath,amssymb,graphicx,multirow,xspace}
\usepackage[dvipsnames]{xcolor}
\usepackage[colorlinks=true,linktocpage=true]{hyperref}
\hypersetup{                   %
    colorlinks = true,         % use colors in links
    linkcolor  = MidnightBlue, % internal links (sections, equations, ...)
    urlcolor   = MidnightBlue,     % external links (email, webpage, ...)
    citecolor  = MidnightBlue,  % citations/references
    anchorcolor = MidnightBlue,
    pagecolor = MidnightBlue
}
\usepackage[compress,numbers,sort]{natbib}
\usepackage{braket}
\usepackage{slashed}
\usepackage{physics}
\usepackage{bm}
\usepackage{indentfirst}
\usepackage{appendix}
\usepackage{verbatim}
\usepackage[compat=1.1.0]{tikz-feynman}
\usepackage{graphicx}
\usepackage{cancel}
\usepackage{cleveref}
\crefname{equation}{Eq.}{Eqs.} % capitalize "E", no period
\crefrangelabelformat{equation}{(#3#1#4--#5#2#6)}

\usepackage[T1]{fontenc}
\long\def\symbolfootnote[#1]#2{\begingroup%
\def\thefootnote{\fnsymbol{footnote}}\footnote[#1]{#2}\endgroup}

\newcommand{\gev}{\mathrm{GeV}}
\newcommand{\tev}{\mathrm{TeV}}
\renewcommand{\vev}[1]{\langle #1 \rangle}

\renewenvironment{abstract}{
    \vspace{0.2cm}
    \centerline{\normalfont\large\bfseries Abstract}
    \begin{quote}
    }
    {
    \end{quote}
}

\input prepictex
\input pictex
\input postpictex
\newdimen\tdim
\tdim=\unitlength

\begin{document}
\setcounter{tocdepth}{2}
\begin{titlepage}
    \vspace*{-1.5cm}
    \hfill{FERMILAB-PUB-25-0524-T}
    \vspace{0.2
        cm}
    \begin{center}
        {\LARGE\bf
            Upper Bound on Parity Breaking Scale for \\
            Doublet WIMP Dark Matter    \par}
    \end{center}
    \vspace{0.2cm}

    \begin{center}
        {\large
            Matthew J.~Baldwin,$^{1,2}$\symbolfootnote[1]{mjbaldwin@uchicago.edu}
            Keisuke Harigaya,$^{1,2,3}$\symbolfootnote[2]{kharigaya@uchicago.edu}
            Isaac R. Wang,$^{4,5}$\symbolfootnote[3]{isaacw@fnal.gov}
        }\\
        \vspace{0.5cm}
        \textit{
            $\,^1$ Department of Physics, University of Chicago, Chicago, IL 60637, USA \\
            \vspace{0.5cm}
            $\,^2$ Enrico Fermi Institute, Leinweber Institute for Theoretical Physics, and Kavli Institute for Cosmological Physics, University of Chicago, Chicago, IL 60637, USA \\
            \vspace{0.5cm}
            $\,^3$ Kavli Institute for the Physics and Mathematics of the Universe (WPI),\\
            The University of Tokyo Institutes for Advanced Study,\\
            The University of Tokyo, Kashiwa, Chiba 277-8583, Japan \\
            \vspace{0.5cm}
            $\,^4$ New High Energy Theory Center, \\
            Department of Physics and Astronomy,\\
            Rutgers University, NJ 08854, USA\\
            \vspace{0.5cm}
            $\,^5$ Theory Division,\\
            Fermi National Accelerator Laboratory,\\
            IL 60510, USA}\\
    \end{center}

    \vspace{0.3cm}

    \begin{abstract}
        We consider weakly interacting massive particle (WIMP) dark matter in a Parity solution to the strong CP problem. The WIMP phenomenology can be drastically affected by the presence of Parity partners of the WIMP and electroweak gauge bosons. We focus on a Parity extension of $SU(2)_L$-doublet fermion dark matter, identify the viable parameter space, and derive the predictions of the theory. We find that the Parity symmetry breaking scale is bounded from above, with the bound given by $25-60$ TeV, depending on whether or not dark matter and its Parity partner coannihilate with each other.
        The High-Luminosity Large Hadron Collider, future colliders, and direct and indirect detection experiments can probe the parameter space further, with correlated signal rates.
    \end{abstract}

\end{titlepage}

\vspace{0.2cm}
\noindent

\noindent\makebox[\linewidth]{\rule{\textwidth}{1pt}}
\tableofcontents
\noindent\makebox[\linewidth]{\rule{\textwidth}{1pt}}

\newpage
\section{Introduction}

The symmetry structure of the Standard Model (SM) of particle physics remains mysterious. The weak interaction violates CP symmetry via an $O(1)$ phase in the CKM matrix, while the  strong CP phase should be smaller than $10^{-10}$~\cite{nEDM:2020crw}.
Such an unnatural hierarchy is known as the strong CP problem.
Several ways to understand this hierarchy in the CP violation parameters have been suggested, including Peccei-Quinn symmetry~\cite{Peccei:1977hh,Peccei:1977ur}, CP symmetry~\cite{Nelson:1983zb,Barr:1984qx,Bento:1991ez,Hiller:2001qg,Vecchi:2014hpa,Dine:2015jga,Girmohanta:2022giy,Kuchimanchi:2023imj,Feruglio:2024ytl}, and Parity symmetry~\cite{Beg:1978mt,Mohapatra:1978fy}.

In particular, the solution via Parity symmetry provides multiple theoretical advantages over alternative approaches. Unlike the Peccei-Quinn symmetry, which must have a QCD anomaly, Parity symmetry can be an exact symmetry that is spontaneously broken.
Additionally, in contrast to CP symmetry, Parity symmetry permits CP phases in the Yukawa interactions, and hence
the CKM phase can be readily obtained.

Various realizations of Parity symmetry and its breaking have been developed, among which the minimal extended gauge group is $SU(3)_c\times SU(2)_L\times SU(2)_R\times U(1)_X$.
In this framework, $SU(2)_R\times U(1)_X$ is spontaneously broken to $U(1)_Y$ via the vacuum expectation value (VEV) of a scalar in the $({\bf 1},{\bf 1},{\bf 3},1)$~\cite{Beg:1978mt,Mohapatra:1978fy,Kuchimanchi:1995rp,Mohapatra:1995xd} or $({\bf 1},{\bf 1},{\bf 2},1/2)$~\cite{Babu:1988mw,Babu:1989rb,Hall:2018let} representation, denoted as $v_R$.
The latter case, where the $SU(2)_R\times U(1)_X$ breaking Higgs $H_R$ is the Parity partner of the SM Higgs $H_L$ in $({\bf 1},{\bf 2},{\bf 1},1/2)$, exhibits particular theoretical merits. First, the extra hierarchy problem associated with the Parity breaking scale can be avoided~\cite{Hall:2018let}. Second, no physical phases are present in the scalar sector, and thereby a possible strong CP phase coming from the CP-violating phases in the VEVs of the Higgses is absent.
Consequently, the strong CP problem can be solved without the necessity of introducing extra symmetries beyond Parity.
We refer to such a setup as the \textit{minimal Higgs content} and will primarily focus on it throughout this paper. We impose Parity on the theory up to possible soft Parity breaking mass terms of $H_L$ and $H_R$, which can be understood as a result of the spontaneous breaking of Parity at a scale above $v_R$. We call $v_R$ the \textit{Parity breaking scale}, even with this soft breaking.

Beyond these theoretical considerations, Parity symmetry also  addresses several shortcomings of the SM while providing experimental signatures across multiple research frontiers.
Parity predicts right-handed neutrinos as partners to SM neutrinos, which may be responsible for the generation of non-zero neutrino masses~\cite{Minkowski:1977sc,Yanagida:1979as,Mohapatra:1979ia,GellMann:1980vs}, the observed baryon asymmetry of the universe~\cite{Fukugita:1986hr,Buchmuller:2004nz,Giudice:2003jh}, and dark matter (DM) (see \cite{Babu:1988yq,Dror:2020jzy,Dunsky:2020dhn,Harigaya:2021txz,Babu:2022ikf,Harigaya:2022wzt,Carrasco-Martinez:2023nit,Hall:2023vjb,Carrasco-Martinez:2025zus} for works on models with the minimal Higgs content).
If sufficiently light, so as to behave as dark radiation, right-handed neutrinos contribute to the effective neutrino number $\Delta N_{\rm eff}$, providing a cosmological probe of Parity symmetric theories~\cite{Babu:2022ikf,Harigaya:2021txz,Harigaya:2022wzt}.
If sufficiently heavy, they manifest as heavy neutral leptons that can be directly produced and detected at diverse ongoing accelerator experiments~\cite{Abdullahi:2022jlv,Hall:2023vjb}.
Parity extends the SM gauge group to $SU(3)_c\times SU(2)_L\times SU(2)_R\times U(1)_X$,
predicting Parity partners of electroweak gauge bosons $W_R$ and $Z'$ that are under intensive searches at collider experiments~\cite{ATLAS:2019lsy,ATLAS:2018tvr,ATLAS:2019erb}.
Furthermore, the phase transition of the $SU(2)_R \times U(1)_X$ breaking could potentially provide another opportunity for baryogenesis~\cite{Harigaya:2022wzt}.
DM candidates in Parity symmetric theories have also been studied in the literature,
including right-handed neutrinos mentioned above, mirror electrons~\cite{Dunsky:2019api,Bonnefoy:2023yoj}, and weakly interacting massive particles (WIMP)~\cite{Heeck:2015qra,Garcia-Cely:2015quu,Kawamura:2018kut,Baldwin:2024bob}.

In this paper, we focus on WIMP DM,  a predictive framework where the DM abundance is determined by the freeze-out mechanism~\cite{Lee:1977ua}.
Parity symmetry has profound implications on WIMP DM.
In standard WIMP scenarios, the DM mass scale is predicted to be around the TeV scale if the annihilation of DM proceeds via electroweak interactions.
However, imposing Parity symmetry on the WIMP scenario introduces additional particles in the DM sector, extending the interaction of DM beyond the SM electroweak gauge bosons to include their Parity partners.
These new interactions within the DM sector, as well as the annihilation channels enabled by them, can dramatically alter the prediction on the mass scale of DM and the experimental strategies to probe the DM sector. We note that WIMP models in a Parity symmetric theory with $SU(2)_R\times U(1)_X$ and Parity breaking by a scalar in $({\bf 1},{\bf 1},{\bf 3},1)$ and $SU(2)_L\times U(1)_Y$ breaking by a scalar in $({\bf 1},{\bf 2},{\bf 2},0)$ are surveyed in~\cite{Heeck:2015qra,Garcia-Cely:2015quu}.%
\footnote{Refs.~\cite{Heeck:2015qra,Garcia-Cely:2015quu} discuss a left-right symmetric model without space-time Parity, but whether or not space-time Parity is involved does not affect the WIMP phenomenology.}

We investigate the Parity extension of fermion WIMP DM in the minimal electroweak representation, namely, $({\bf 2},-1/2)$ of $SU(2)_L \times U(1)_Y$.
The mass of fermion DM is protected by chiral symmetry, and there is no extra hierarchy problem associated with the mass of DM.
In the minimal $SU(3)_c \times SU(2)_L \times SU(2)_R \times U(1)_X$ group,
$({\bf 2},-1/2)$ DM can be embedded into  $({\bf 2}, {\bf 2},0)$ or $({\bf 2}, {\bf 1},-1/2)\oplus({\bf 1}, {\bf 2},-1/2)$ of $SU(2)_L\times SU(2)_R\times U(1)_X$.
The former reduces to ordinary Dirac fermion DM in $({\bf 2},-1/2)$ of $SU(2)_L \times U(1)_Y$ and is excluded by DM direct-detection experiments unless the model is extended to introduce a Majorana mass term; see Appendix~\ref{app:bidoublet} for the analysis of the model. The latter is a viable scenario that we focus on in the main part of the paper.

We find that the Parity and $SU(2)_R\times U(1)_X$ symmetry breaking scale $v_R$ is bounded from {\it both above and below} by DM physics.
The electrically neutral component in $({\bf 1}, {\bf 2},-1/2)$ serves as DM and is a singlet under the SM gauge group.
When $s$-channel resonant annihilation via $W_R$ or $Z'$ is not effective, the freeze-out mechanism predicts the mass of DM and its partners to be around 260 GeV.
A hyper-charged particle in the DM sector is predicted to be long-lived, and the model is excluded by charged-track searches at the Large Hadron Collider (LHC)~\cite{ATLAS:2019gqq}.
Consequently, the DM mass must be around $m_{W_R}/2$ or $m_{Z'}/2$ to annihilate via resonance.
If $v_R$ and the DM mass are sufficiently large,
DM annihilation is suppressed, even at the center of the resonance, and DM is overproduced, providing an upper bound on $v_R$.
Scattering with nucleons via $Z'$ exchange provides a lower bound on $v_R$ via direct detection, which complements the upper bound from DM overproduction.
The parameter space of the theory can be further probed by the High-Luminosity LHC (HL-LHC) through new gauge boson searches and future direct and indirect detection experiments, with their signals correlated with each other.

The upper bound on $v_R$ is encouraging also for new-physics searches other than DM searches. Without the bound, $v_R$ can be anywhere between $10$ TeV and a very high energy scale bounded only by corrections to the strong CP phase by higher dimensional operators (see Sec.~\ref{sec:model}), and it is not clear if new-physics signals from Parity are within the reach of near-future experiments. With the upper bound from the DM physics, there will be a good chance of discovery of new physics in near future.

This paper is organized as follows.
In Sec.~\ref{sec:model}, after a short review of the Parity solution to the strong CP problem, we present a doublet WIMP model.
In Sec.~\ref{sec:relic}, we calculate the relic abundance of WIMP DM.
In Sec.~\ref{sec:signal}, we discuss various experimental probes of the model, including collider searches, direct detection, and indirect detection.
Sec.~\ref{sec:summary} summarizes our results and conclusions.

\section{Parity symmetric WIMP model}
\label{sec:model}

In this section, we review a Parity symmetric extension of the SM and discuss the embedding of $SU(2)$ doublet WIMP DM into Parity symmetric theories.

\subsection{Parity extension of the SM}
We first discuss how the SM can be embedded into Parity symmetric theories.
The gauge symmetry at the UV scale is $SU(3)_c\times SU(2)_L\times SU(2)_R\times U(1)_X$.
This symmetry breaks in two stages.
First, $SU(2)_R\times U(1)_X$ is broken down to the SM $U(1)_Y$ at a scale much above the EW scale, with hypercharge $Y= T_{3R}+X$.
Second, electroweak symmetry breaking (EWSB) occurs, which breaks $SU(2)_L \times U(1)_Y$ down to $U(1)_{\rm EM}$ at the EW scale, with electromagnetic (EM) charge $Q= T_{3L}+Y$.
We consider a minimal Higgs model~\cite{Babu:1988mw,Babu:1989rb,Hall:2018let} whose Higgs and (a part of) fermion contents are shown in Table~\ref{tab:minimal Higgs}.
The $SU(2)_R$ breaking is driven by $H_R({\bf 1},{\bf 1},{\bf 2},-1/2)$ while the $SU(2)_L$ breaking is driven by $H_L({\bf 1},{\bf 2},{\bf 1},1/2)$.

\begin{table}[tbp]
    \caption{The gauge charges of Higgses and fermions}
    \begin{center}
        \begin{tabular}{|c|c|c|c|c|c|c|} \hline
                      & $H_L$         & $H_R$          & $q_i$         & $\bar{q}_i$     & $\ell_i$        & $\bar{\ell}_i$ \\ \hline
            $SU(3)_c$ & {\bf 1}       & {\bf 1}        & {\bf 3}       & ${\bf \bar{3}}$ & {\bf 1}         & {\bf 1}        \\
            $SU(2)_L$ & {\bf 2}       & {\bf 1}        & {\bf 2}       & {\bf 1}         & {\bf 2}         & {\bf 1}        \\
            $SU(2)_R$ & {\bf 1}       & {\bf 2}        & {\bf 1}       & {\bf 2}         & {\bf 1}         & {\bf 2}        \\
            $U(1)_X$  & $\frac{1}{2}$ & $-\frac{1}{2}$ & $\frac{1}{6}$ & $-\frac{1}{6}$  & $- \frac{1}{2}$ & $\frac{1}{2}$  \\ \hline
        \end{tabular}
    \end{center}
    \label{tab:minimal Higgs}
\end{table}%

We begin by examining how the two symmetry breakings generate gauge boson masses through gauge interactions.
The scalar fields $H_L$ and $H_R$ acquire VEVs as
\begin{equation}
    \label{eq:vev}
    \vev{H_L} = \begin{pmatrix}
        0 \\ v_L
    \end{pmatrix},~~
    \vev{H_R} = \begin{pmatrix}
        v_R \\ 0
    \end{pmatrix}.
\end{equation}
We use the convention where $v_L \simeq 174~\gev$.
The kinetic terms are
\begin{align}
    \mathcal{L} = |D_\mu H_L|^2 + |D_\mu H_R|^2.
\end{align}
Here, the covariant derivatives are
\begin{align}
    D_\mu H_L  & = \left(  \partial_\mu - i g_L W_{L \mu} - i \frac{1}{2}g_X B_{X\mu} \right) H_L, \nonumber \\
    D_\mu H_R  & = \left(  \partial_\mu - i g_R W_{R \mu} + i \frac{1}{2}g_X B_{X\mu} \right) H_R, \nonumber \\
    W_{L,R\mu} & \equiv \frac{1}{2}\sigma_i W_{L,R\mu}^i.
\end{align}
Although Parity enforces $g_L = g_R$ above $v_R$, for now, we distinguish $g_L$ from $g_R$.
We assume $g_L = g_R = g$ when calculating the Higgs-gauge boson interactions and annihilation cross sections of DM and define $\alpha_2 \equiv g^2/(4\pi)$.
Similar to the SM EW sector, the massless $U(1)_Y$ gauge boson $B_Y$ is a mixture of  $W^3_{R}$ and $B_X$.
Defining $\tan \theta_{R} \equiv g_X/g_R$, we have the following gauge boson mixing,
\begin{align}
    B_{X\mu}   & = \cos \theta_{R} B_{Y\mu} - \sin \theta_{R} Z_\mu', \nonumber \\
    W_{R\mu}^3 & = \sin \theta_{R} B_{Y\mu} + \cos \theta_{R} Z_\mu'.
\end{align}
The $Z'$ and $W_R$ gauge bosons acquire masses from $v_R$.
This symmetry breaking relates
\begin{align}
    \frac{1}{g_Y^2} = \frac{1}{g_R^2} + \frac{1}{g_X^2},
\end{align}
similar to the SM EWSB. $SU(2)_L \times U(1)_Y$ is spontaneously broken down to $U(1)_{\rm EM}$ via EWSB. Defining $\tan \theta_L \equiv g_Y/g_L$, we have the following mixing between the $B_Y$ and $W_L^3$ gauge bosons,
\begin{align}
    B_{Y\mu}   & = \cos \theta_{L} A_\mu - \sin \theta_{L} Z_\mu \nonumber \\
    W_{L\mu}^3 & =  \sin \theta_{L} A_\mu + \cos \theta_{L} Z_\mu.
\end{align}
Hereafter we will denote $s_{L,R}\equiv \sin\theta_{L,R}$, $c_{L,R}\equiv \cos\theta_{L,R}$ and $t_{L,R}\equiv \tan\theta_{L,R}$.

The above two symmetry breakings generate the following gauge boson masses,
\begin{align}
    \label{eq:gauge boson mass}
    m_{Z'}^2 = \frac{1}{2}(g_R^2 + g_X^2) v_R^2,\ m_{W_R}^2 = \frac{1}{2} g_R^2 v_R^2,\ m_Z^2 = \frac{1}{2} (g_L^2 + g_Y^2) v_L^2,\ m_{W_L}^2 = \frac{1}{2}g_L^2 v_L^2.
\end{align}
The interactions of the SM Higgs $h_L$ with the gauge bosons are given in Appendix~\ref{app:gauge_interactions}.

With only the fermion content shown in Table~\ref{tab:minimal Higgs}, one cannot write down renormalizable Yukawa interactions. To obtain non-zero SM fermion masses, we introduce several pairs of Dirac fermions. Their Yukawa interactions are
\begin{align}
    {\cal L} = - x_{ij} f_i \bar{X}_j H_L^{(*)}  - x_{ij}^* \bar{f}_i X_j H_R^{(*)} - M_{ij}X_i \bar{X}_j +  {\rm h.c.},
    \label{eq:type1}
\end{align}
\begin{center}or \end{center}
\begin{align}
    {\cal L} = - x_{ij} f_i \bar{X}_j H_R^{(*)}  - x_{ij}^* \bar{f}_i X_j H_L^{(*)} - M_{ij}X_i \bar{X}_j +  {\rm h.c.},
    \label{eq:type2}
\end{align}
where $X$ and $\bar{X}$ are additional Weyl fermions,
$f$ represents $q$ or $\ell$, and $\bar{f}$ represents $\bar{q}$ or $\bar{\ell}$.
There are various possibilities for the gauge charges of $X$ and $\bar{X}$, which can be found in~\cite{Hall:2018let}. For example, to obtain the down-like Yukawas we may consider $D({\bf 3 }, {\bf 1 }, {\bf 1 },-1/3)$ and $\bar{D}({\bf \bar{3} }, {\bf 1 }, {\bf 1 },1/3)$ with couplings and masses
\begin{align}
    {\cal L} = - x_{D} q \bar{D} H_L^{*}  - x_{D}^* \bar{q} D H_R^{*} - m_{D} D \bar{D} +  {\rm h.c.}.
\end{align}
Here and hereafter, we suppress the generation indices.
Generically, the SM $SU(2)_L$ doublet fermions are linear combinations of $f$ and $X$ and the $SU(2)_L$ singlet fermions are those of $\bar{f}$ and $\bar{X}$.
When $M \gg x v_R$, we can integrate out the Dirac fermions to obtain dimension-5 operators $x^2 H_L^{(*)} H_R ^{(*)} f \bar{f} / M $ and the SM fermion mass is given by $x^2v_Lv_R/M$.
In the opposite limit, the non-zero VEV of $H_R$ gives masses of $x v_R$ to the fermions that it directly couples to, and the Parity partners of these fermions obtain masses of $x v_L$ from their coupling with $H_L$.

The strong CP problem is solved by the above Lagrangian with Parity symmetry.
The Dirac mass term $M_{ij}$ is Hermitian as required by Parity.
It follows that in both Eqs.~\eqref{eq:type1} and \eqref{eq:type2}, the determinant of the quark mass matrix is real, and hence the strong CP problem is solved. Strong CP phases are generated via quantum corrections, but these have been shown to be below the current upper bound~\cite{Hall:2018let,Hisano:2023izx}.
Unlike the cases where the SM Higgs is embedded into an $SU(2)_L\times SU(2)_R$ bi-fundamental, the VEVs of the Higgs fields do not have any physical phases and hence do not generate an additional strong CP phase at tree level.

Higher-dimensional operators can also induce a strong CP phase. The dimension-6 operator
\begin{align}
    \frac{1}{8\pi^2} \frac{|H_L|^2- |H_R|^2}{\Lambda^2} G \tilde{G}
\end{align}
is Parity symmetric and not controlled by chiral symmetry,%
\footnote{
    It can be suppressed by CP symmetry or supersymmetry~\cite{Hall:2018let} broken at higher energy scales.
}
where $\Lambda$ is a cutoff scale.
This operator gives a strong CP phase of $v_R^2/\Lambda^2$, which is small enough if
\begin{equation}
    v_R < 10^{13}{\rm GeV} \frac{\Lambda}{10^{18} {\rm GeV}}.
\end{equation}
In our paper we consider $v_R= 10-100$ TeV, for which the correction to the strong CP phase is small unless the cutoff scale is much below the Planck scale.

We now turn to the scalar potential that generates proper values of $v_L$ and $v_R$.
To satisfy the experimental lower bounds on new gauge boson masses, we need $v_R \gtrsim 14$ TeV $\gg v_L$. The hierarchy in the VEVs can be obtained in two ways. One way, proposed in~\cite{Hall:2018let}, takes the potential to be
\begin{equation}
    V(H_L,H_R)= \lambda \left(|H_L|^2 + |H_R|^2 - v_R^2\right)^2 + \Delta \lambda|H_L^2||H_R|^2  + V_{\rm CW},
\end{equation}
where $|\Delta \lambda| = O(1/16\pi^2)$ and $V_{\rm CW}$ is the Coleman-Weinberg potential. The first term in the potential fixes $H_L$ and $H_R$ on the rotational symmetric space $|H_L|^2 + |H_R|^2 = v_R^2$, and the second and third terms fix $H_L$ and $H_R$ on $v_L \ll v_R$. In this mechanism, the SM Higgs is a pseudo-Nambu Goldstone boson with a global symmetry breaking scale at $v_R$, so the SM Higgs quartic coupling vanishes at $v_R$. If the running of the quartic coupling is given by the SM one, $v_R$ is predicted to be in the range $10^{9-14}$ GeV, depending on the values of the top quark Yukawa and the strong coupling constant. If there exist extra Yukawa interactions of the SM Higgs below the scale $v_R$, then $v_R$ can be lower.
Alternatively, the hierarchy can be achieved by introducing a Parity breaking sector with a Parity-odd order parameter ${\cal O}$ and a coupling
\begin{equation}
    {\cal O} \left( |H_L|^2 - |H_R|^2\right).
    \label{eq:parity breaking}
\end{equation}
After the Parity breaking by $\vev{{\cal O}} \neq 0$, the mass of $H_L$ can be smaller than $H_R$, so that $v_L \ll v_R$. Simple examples of Parity breaking sectors include pure Yang-Mills theories with $\theta = \pi$~\cite{Witten:1980sp,tHooft:1981bkw,Witten:1998uka,Gaiotto:2017yup,Kitano:2020mfk}.
Even in the latter setup, we call $v_R$ the Parity symmetry breaking scale in the following, as it is the parameter determining the low energy phenomenology of Parity symmetric theories.

In both mechanisms, despite the existence of the intermediate scale $v_R$, the total fine-tuning to obtain $v_L < v_R < \Lambda$ is the same as that of the SM,  where $\Lambda$ is the cut-off scale of the theory.
$v_R<\Lambda $ is obtained by tuning the mass term $-2\lambda v_R^2(|H_L|^2 + |H_R|^2)$ with an accuracy of $v_R^2/\Lambda^2$. $v_L < v_R$ is obtained by tuning $\Delta \lambda$ or the coupling in Eq.~\eqref{eq:parity breaking} with an accuracy of $v_L^2/v_R^2$. The total fine-tuning is $v_L^2/\Lambda^2$, which is the same as that of the SM with a cutoff scale $\Lambda$.
The tuning may be explained by environmental selection~\cite{Agrawal:1997gf,Hall:2014dfa,DAmico:2019hih}.

\subsection{Doublet pair WIMP dark matter}\label{sec:WIMPindoubletPair}

The minimal representation of electroweak-charged WIMP DM is $({\bf 2},1/2)\oplus({\bf 2},-1/2)$ of $SU(2)_L\times U(1)_Y$, which combine to form a Dirac particle.
We consider the minimal Parity symmetric model in which the WIMP DM is embedded into two pairs of vector-like doublets that are the Parity partners of each other, $({\bf 2},{\bf 1},1/2)$ $\oplus$ $({\bf 2},{\bf 1},-1/2)$ $\oplus$ $({\bf 1},{\bf 2},1/2)$ $\oplus$ $({\bf 1},{\bf 2},-1/2)$.
We discuss an embedding into $({\bf 2},{\bf 2},0)$ in Appendix~\ref{app:bidoublet}.

To ensure DM stability, we impose a $Z_2$ symmetry and assign an odd charge to the DM multiplets.
We embed the first two and the last two into Dirac fields that we call $\psi_\ell$ and $\psi_r$, respectively. They are in the $({\bf 2},{\bf 1},-1/2)$ and $({\bf 1},{\bf 2},-1/2)$ representations of $SU(2)_L\times SU(2)_R\times U(1)_X$.
These Dirac fields are decomposed as
\begin{equation}
    \psi_\ell = \begin{pmatrix}
        \psi_\ell^0 \\
        \psi_\ell^-
    \end{pmatrix},\
    \psi_r = \begin{pmatrix}
        \psi^0_r \\  \psi_r^-
    \end{pmatrix},~~
\end{equation}
where the superscripts denote their EM charges.
The gauge interactions of $\psi_\ell$ and $\psi_r$ are shown in Appendix~\ref{app:gauge_int_psilr}. We will refer to this model as the doublet pair WIMP DM model.

Note that the DM multiplets cannot have Yukawa interactions with the Higgs fields because of the gauge symmetry, which makes the model predictable. This is in contrast to a class of models with scalars in $({\bf 3},{\bf 1}, 1)$ and $({\bf 1},{\bf 3}, 1)$, in which these scalars can have Yukawa interactions with the DM multiplets to affect the DM phenomenology through generating mass splitting and providing extra annihilation channels.
There can be, however, higher-dimensional interactions such as $\frac{1}{\Lambda}\psi_\ell H_L \psi_r H_R^\dagger$. Although they do not affect DM annihilation as long as $\Lambda$ is much larger than the DM mass, they lead to the decay of $\psi_\ell$ into $\psi_r + h_L/Z/W_L$. The decay plays an important role in the DM phenomenology, as we will see.

After the $SU(2)_R\times U(1)_X$ breaking into $U(1)_Y$, $\psi_\ell$, $\psi_r^-$ and $\psi_r^0$ obtain different masses by quantum corrections,
\begin{align}
    \label{eq:mass splitting pair doublet}
    m_{\psi_r^-} - m_{\psi_{r}^0}  & = \frac{\alpha _2 m_\psi s_R^2 }{4 \pi }f\left(\frac{m_{Z'}^2}{m_\psi^2}\right) , \nonumber                                                                      \\
    m_{\psi_\ell} - m_{\psi_{r}^0} & = \frac{\alpha _2 m_\psi}{8 \pi }\left(\left(1-\frac{1}{2} c_R^2\right) f\left(\frac{m_{Z'}^2}{m_\psi^2}\right)+f\left(\frac{m_{W_R}^2}{m_\psi^2}\right)\right),
\end{align}
where
\begin{align}
    \label{eq:loop_f}
    f(z)\equiv\frac{1}{2} z^2 \log (z)-z-\sqrt{z(z-4)} (z+2)\log \left(\frac{1}{2} \left(\sqrt{z-4}+\sqrt{z}\right)\right).
\end{align}
The mass splitting is shown in Fig.~\ref{fig:mass splitting doublet}. Details of the quantum correction calculations can be found in Appendix~\ref{sec:quantum correction}.
The lightest gauge eigenstate, $\psi_r^0$, becomes the DM candidate.

\begin{figure}[!t]
    \centering
    \includegraphics[width=0.48\linewidth]{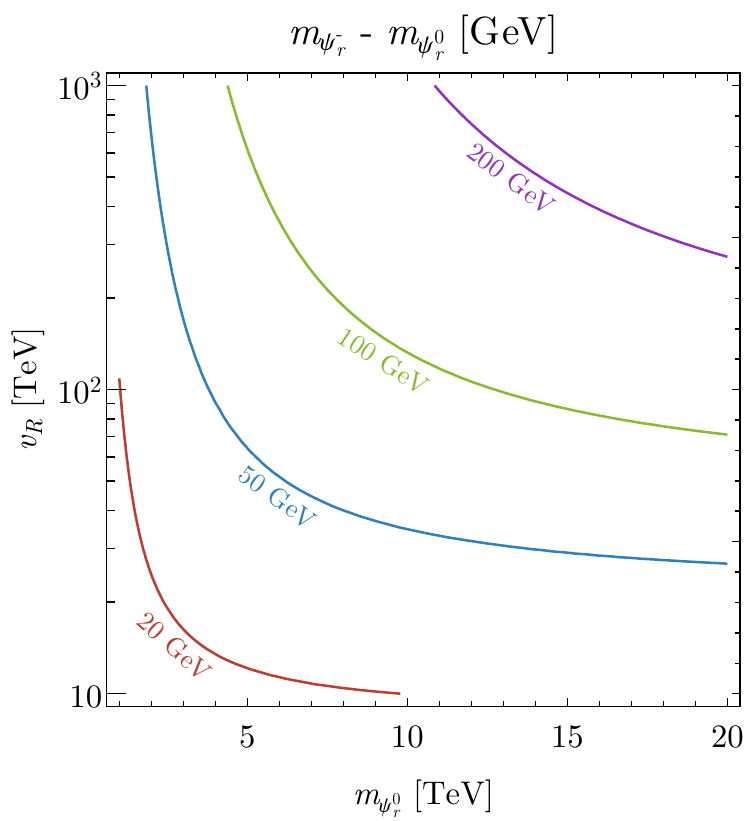}
    \includegraphics[width=0.48\linewidth]{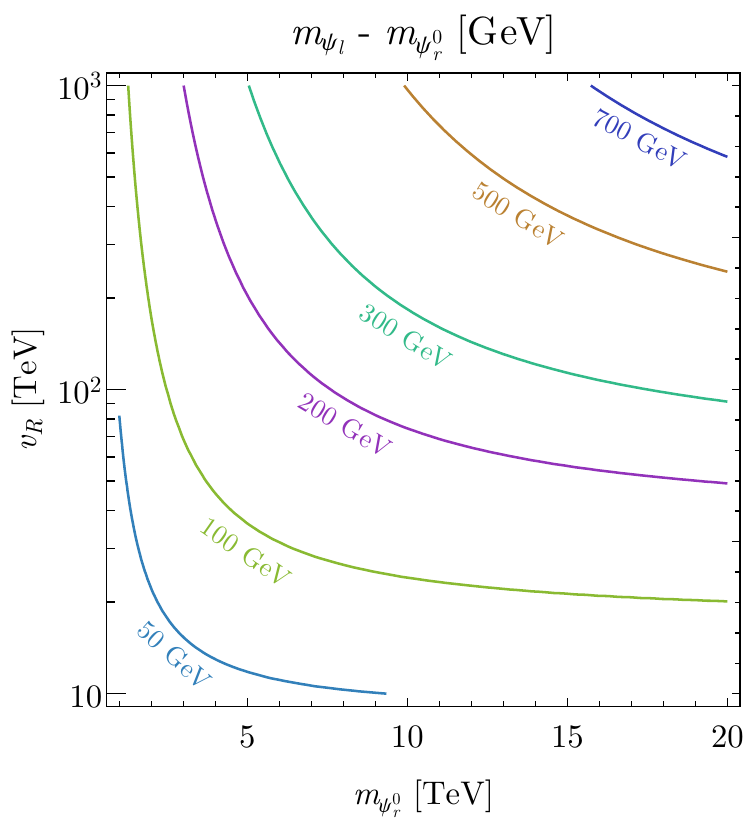}
    \caption{Mass splitting between particles in the DM sector generated by quantum corrections from gauge interactions.}
    \label{fig:mass splitting doublet}
\end{figure}

Furthermore, below the electroweak scale,
$SU(2)_L\times U(1)_Y$ symmetry is broken and $\psi_\ell$ breaks into $\psi_\ell^0$ and $\psi_\ell^-$.
Quantum corrections generate a small mass splitting between them~\cite{Cirelli:2005uq,Ibe:2012sx},
\begin{equation}
    m_{\psi_\ell^-} - m_{\psi_{\ell}^0} = \frac{\alpha _2 m_\psi s_L^2 }{4 \pi }f\left(\frac{m_{Z}^2}{m_\psi^2} \right)\,,
\end{equation}
with $m_{\psi_\ell^0} = m_{\psi_\ell}$ computed from Eq.~\eqref{eq:mass splitting pair doublet}.

\newpage
\section{Thermal relic abundance}
\label{sec:relic}
In this section, we compute the thermal relic abundance of DM.

\subsection{Freeze-out mechanism and coannihilation}

The relic abundance of WIMP DM is determined by thermal freeze-out.
Initially, WIMP DM remains in thermal equilibrium throughout  production and annihilation processes.
As the temperature drops below the DM mass, DM annihilates into other particles, and the number density of DM becomes exponentially suppressed.
As the universe continues to expand and cool, the annihilation rate eventually falls below the Hubble expansion rate and the DM number density freezes out.

The DM relic abundance is computed by solving the Boltzmann equation.
When DM and other exotic particles can be converted into each other, they annihilate together, which is called coannihilation.
The Boltzmann equation for DM including coannihilation is~\cite{Griest:1990kh}
\begin{align}
    \label{eq:Boltzmann}
    \frac{dY}{dx} = - \frac{\langle \sigma v \rangle_{\rm eff}}{H x} \left( 1 - \frac{x}{3 g_{*s}} \frac{d g_{*s}}{dx} \right)s \left( Y^2 - Y_{\rm eq}^2 \right).
\end{align}
Here we have defined the time variable $x \equiv m_{\psi}/T$, where $m_{\psi}$ is the DM mass.
The entropy density is then expressed as $s \equiv g_{*s} (2 \pi^2/45) (m_{\psi}/x)^3$.
The yield of DM is defined as $Y \equiv n/s$.
The Hubble rate and the equilibrium Boltzmann distribution are given by
\begin{align}
    H (x)          = \sqrt{\frac{g_*}{90}} \frac{\pi}{M_{\rm pl}} \left( \frac{m_{\psi}}{x} \right)^2,~~
    Y_{\rm eq}(x)  = g_{\rm eff}(x) \frac{m_{\psi}^3}{s} \frac{1}{(2 \pi x)^{3/2}}e^{-x}.
\end{align}
The effective thermally-averaged cross section $\langle \sigma v \rangle_{\rm eff}$  is defined as
\begin{align}
    \label{eq:sigmav_eff}
    \langle \sigma v \rangle_{\rm eff}(x) \equiv \frac{1}{g_{\rm eff}^2} \sum_{ij} \langle \sigma v \rangle_{ij} g_i(x) g_j(x),
\end{align}
with the temperature-dependent degrees of freedom and effective total degrees of freedom defined as
\begin{align}
    \label{eq:dof}
    g_i(x)         & \equiv g_i (1+\Delta_i)^{3/2} \exp(-x \Delta_i), \nonumber \\
    g_{\rm eff}(x) & = \sum_i g_i(x),
\end{align}
where $\Delta \equiv m_i - m_\psi$ is the mass splitting, and $g_i$ is the intrinsic degrees of freedom of each particle species $i$.
For a Majorana fermion, $g_i = 2$. For a Dirac fermion, $g_i = 4$.
The summation runs over the annihilating particle species whose conversion rate with the DM particle is far above the Hubble expansion rate at around the time of freeze-out.
$\langle \sigma v \rangle_{ij}$ is the thermally-averaged annihilation cross section  with initial particle species $i$ and $j$, and is computed in the following subsection.

\subsection{Sommerfeld effect and annihilation cross sections}
\label{sec:annihilation}

The key ingredient of Eq.~\eqref{eq:Boltzmann} is the annihilation cross section.
Since freeze-out occurs at around $m_\psi/T \simeq 20$, we work in the non-relativistic (NR) limit,
where the cross section is dominated by $s$-wave contributions.
For analytical cross section calculations, we define the center-of-mass (CoM) frame assuming both the initial state particles have the same mass $m_\psi$, since the mass difference between new fermions is at most $3\%$ of $m_{\psi_r^0}$.
The value of $m_\psi$ depends on the annihilation process: $m_{\psi_r^0}$ for $\bar{\psi}_r^0 \psi_r^0$ annihilation, $(m_{\psi_r^-} + m_{\psi_r^0})/2$ for $\bar{\psi}_r^0 \psi_r^-$ annihilation, $m_{\psi_r^-}$ for $\bar{\psi}_r^- \psi_r^-$ annihilation, and so forth.
Near the $s$-channel resonance peak, the velocity dependence of $s$ becomes important.
We take $s = 4m_\psi^2$ for most cases, but use $s = 4m_\psi^2 + m_\psi^2 v^2$ in denominators of $s$-channel mediators to properly account for potential $s$-channel resonance effects\footnote{We use $v$ to represent the relative velocity in the initial state. Some papers in the literature instead use $v$ for the velocity of one particle in the CoM frame.}.
Cross sections of all possible annihilation channels are computed in Appendix~\ref{sec:channel}.

When the initial-state fermions are much heavier than the boson mediators, the long-range potential induced by the bosons can significantly alter the wave function of incoming fermions, and thus change the annihilation cross section.
This non-perturbative effect is known as the Sommerfeld effect~\cite{Sommerfeld:1931qaf}, whose impact on DM annihilation has been studied in the literature~\cite{Hisano:2004ds,Hisano:2006nn,Arkani-Hamed:2008hhe,Cassel:2009wt,Slatyer:2009vg,Feng:2010zp}.
A general review is provided here.

The initial-state wave function is governed by the Schr\"{o}dinger equation.
Consider $N$ incoming states $a = 1,2,3,...,N$ with identical total charge and spin, each containing two particles that annihilate.
The two-body reduced wave function $g(r)$ in an $s$-wave satisfies
\begin{align}
    \label{eq:Schroedinger}
    - \frac{1}{m_\psi} \frac{d^2}{dr^2} \mathbf{g}(r) + \mathbf{V}(r) \cdot \mathbf{g}(r) = E \mathbf{g}(r)\,,
\end{align}
where $r$ is the radial coordinate and $E$ is the total energy of the two-body state.
At large distances, the potential vanishes, and the energy approaches the NR kinetic energy, $E \to m_\psi v^2/4$.
Both the wave function $\mathbf{g}(r)$ and the long-range potential $\mathbf{V}(r)$ are $N\times N$ matrices.
We provide the long-range potentials for each case in Appendix~\ref{sec:channel}.

The solution requires two boundary conditions.
First, only outgoing waves survive at infinity,
\begin{align}
    \mathbf{g}_{ab}|_{r\to \infty} \to \sum_k\delta_{a k} d_{k b}(v) \exp (i \frac{m_\psi v r}{2})\,.
\end{align}
Second, the wave function must be finite for all $r$.
The coefficients $d(v)$ can be computed by solving Eq.~\eqref{eq:Schroedinger} numerically with these two boundary conditions.

Such a wave function adds a factor to the annihilation cross section for the initial states as
\begin{align}
    \label{eq:sigmav}
    (\sigma v)_a (v)= c_a \sum_{k,l} \Gamma_{kl} d_{a k}(v) d_{a l}^*(v).
\end{align}
The coefficient $c_a = 2$ for identical particles in the initial state and $c_a = 1$ otherwise.
The absorptive matrix $\Gamma$ has the form
\begin{align}
    \Gamma_{kl}|_{k=l}      & = \frac{1}{c_k} (\sigma v)_k \,,\nonumber                    \\
    \Gamma_{kl}|_{k \neq l} & = \frac{1}{\sqrt{c_k c_l}} (\Gamma_{k \leftrightarrow l})\,.
\end{align}
Here, the diagonal part of this absorptive matrix is the annihilation cross section, while the off-diagonal parts are the transition rates between different initial states.
They are computed in Appendix~\ref{sec:channel}.
Compared to the tree-level result, the extra factors $d_{ak}d_{al}^*$ arising from the long-range potential modify the annihilation cross sections.
This modification is known as the Sommerfeld effect.

Eq.~\eqref{eq:sigmav} is velocity-dependent and should be thermally averaged before entering Eq.~\eqref{eq:sigmav_eff}.
The thermally-averaged cross section is
\begin{align}
    \langle \sigma v \rangle = \left( \frac{x}{4 \pi} \right)^{3/2} \int dv (\sigma v) 4 \pi v^2 \exp(- \frac{x v^2}{4})\,.
\end{align}
With this computed as a function of $x$, we can solve Eq.~\eqref{eq:Boltzmann} to obtain the DM relic abundance.

\subsection{Relic density of doublet pair WIMP dark matter}
\label{sec:relic pair doublet}

Now we apply the above framework to our doublet pair WIMP DM model.
Within each doublet, their components coannihilate: $\psi_\ell$ via $W_L$ exchange and $\psi_r$ via $W_R$ exchange.
$\psi_\ell$ and $\psi_r$, on the other hand, do not coannihilate with each other through renormalizable interactions.
Instead, they would annihilate independently, each acquiring a relic abundance.
If $\psi_\ell^0$ remains stable and contributes to DM, the model is excluded by direct detection experiments due to its vector-current interaction with nucleons.

Additional interactions are thus required to enable $\psi_\ell^0$ decay, which also determines whether $\psi_\ell$ and $\psi_r$ coannihilate.
As a minimal example, $\psi_\ell$ and $\psi_r$ can convert into each other via a higher-dimensional operator $\frac{1}{\Lambda}\psi_\ell \psi_r H_L H_R^*$.
With this operator, $\psi_\ell$ can decay into $\psi_r$ at a rate
\begin{equation}
    \Gamma \simeq
    \begin{cases}
        \frac{1}{4\pi} \frac{v_R^2}{\Lambda^2} (m_{\psi_\ell} - m_{\psi_r})              & :  m_{\psi_\ell} - m_{\psi_r} > m_Z, \\
        \frac{1}{128\pi^3} \frac{v_R^2 (m_{\psi_\ell} - m_{\psi_r})^3 }{\Lambda^2 v_L^2} & : m_{\psi_\ell} - m_{\psi_r} < m_Z.
    \end{cases}
\end{equation}
If the decay is effective at the time of freeze-out, which occurs when
\begin{align}
    \Lambda <
    \begin{cases}
        10^{11}~\mathrm{GeV} \left( \frac{m_{\psi_\ell} - m_{\psi_r^0}}{200~\mathrm{GeV}} \right)^{1/2} \left( \frac{v_R}{20 \mathrm{TeV}} \right) \left( \frac{x}{20} \right) \left( \frac{6 \mathrm{TeV}}{m_\psi} \right) & : m_{\psi_\ell} - m_{\psi_r} > m_Z, \\
        10^{9}~\mathrm{GeV} \left( \frac{m_{\psi_\ell} - m_{\psi_r^0}}{50~\mathrm{GeV}} \right)^{3/2} \left( \frac{v_R}{20 \mathrm{TeV}} \right) \left( \frac{x}{20} \right) \left( \frac{6 \mathrm{TeV}}{m_\psi} \right)   & : m_{\psi_\ell} - m_{\psi_r} < m_Z,
    \end{cases}
\end{align}
$\psi_\ell$ and $\psi_r$ coannihilate.
Otherwise, $\psi_\ell$ and $\psi_r$ annihilate separately.
In this paper, we discuss two extreme cases:
\begin{itemize}
    \item Coannihilation case: decay is effective during freeze-out, allowing $\psi_\ell^0$, $\psi_\ell^-$, $\psi_r^0$, and $\psi_r^-$ to coannihilate with each other. The DM relic abundance is $\Omega_{\rm DM}h^2 = \Omega_{\psi_r^0}h^2$.
    \item Non-coannihilation case:
          Decay occurs much after freeze-out.
          $\psi_r^0$ and $\psi_r^-$ coannihilate to set $\Omega_{ \psi_r^0}h^2$, while $\psi_\ell^0$ and $\psi_\ell^-$ coannihilate to set $\Omega_{\psi_\ell^0} h^2$. $\psi_\ell^0$ later decays into $\psi_r^0$, so that the DM relic abundance is given by $\Omega_{\rm DM}h^2 = \Omega_{\psi_r^0} h^2 + m_{\psi_r^0} \Omega_{\psi_\ell^0}h^2/m_{\psi_\ell^0}$.
\end{itemize}

\begin{figure}[!t]
    \centering
    \includegraphics[width=0.49\linewidth]{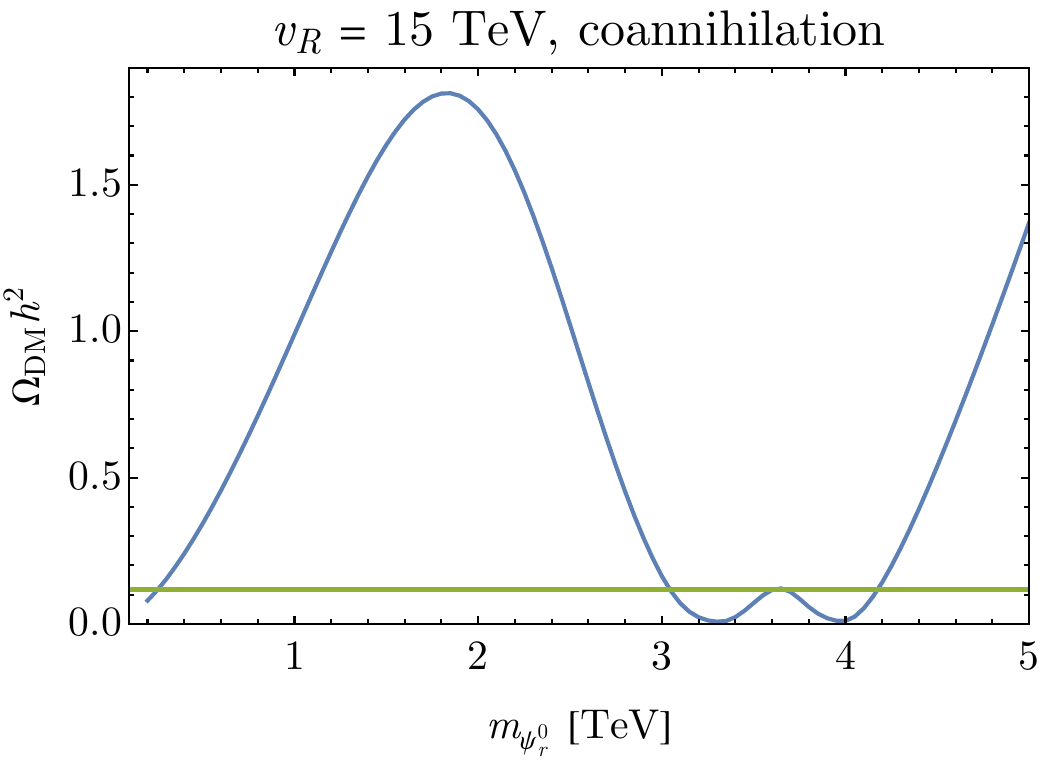}
    \includegraphics[width=0.49\linewidth]{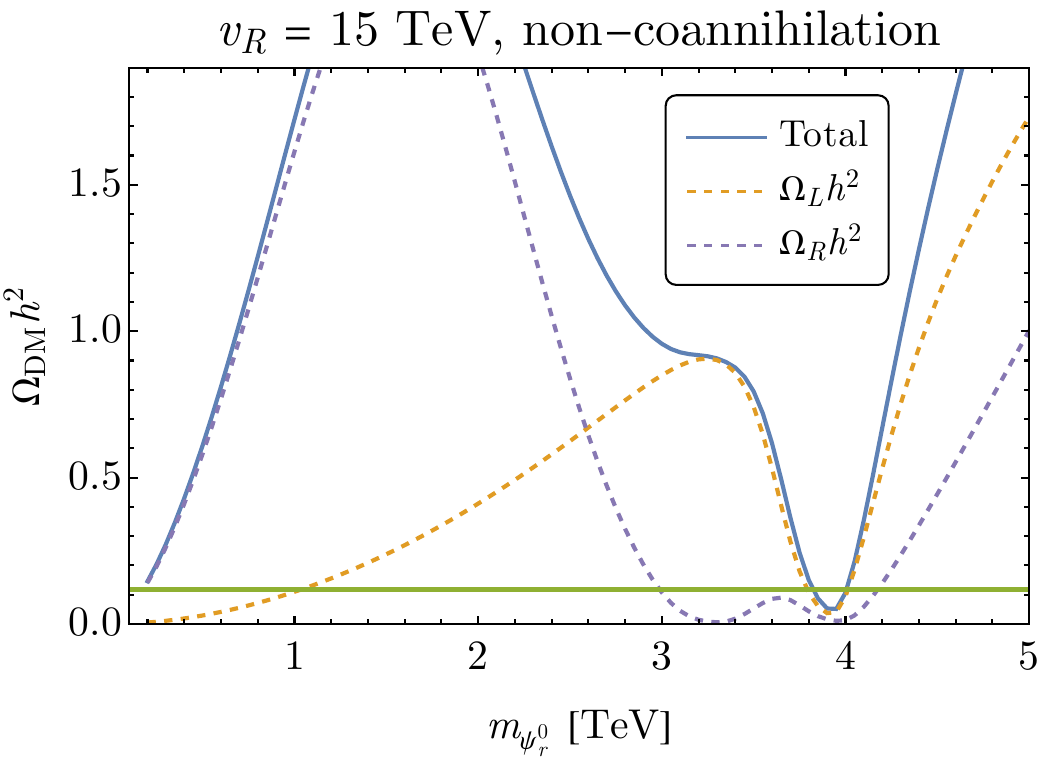}
    \caption{The DM relic density as a function of the DM mass $m_{\psi_r^0}$ with a fixed $v_R = 15$ TeV. Left: coannihilation case. Right: non-coannihilation case. We show the contributions from $\Omega_R h^2\equiv\Omega_{\psi_r^0}h^2$ and $\Omega_L h^2\equiv m_{\psi_r^0}\Omega_{\psi_\ell^0} h^2/m_{\psi_\ell^0}$ with dashed lines. The green bands corresponds to $\Omega_{\rm DM}h^2 = 0.1186\pm0.0022$.}
    \label{fig:doublet-pair-fixvR}
\end{figure}

\begin{figure}[!t]
    \centering
    \hspace*{-0.5cm}
    \includegraphics[width=0.49\linewidth]{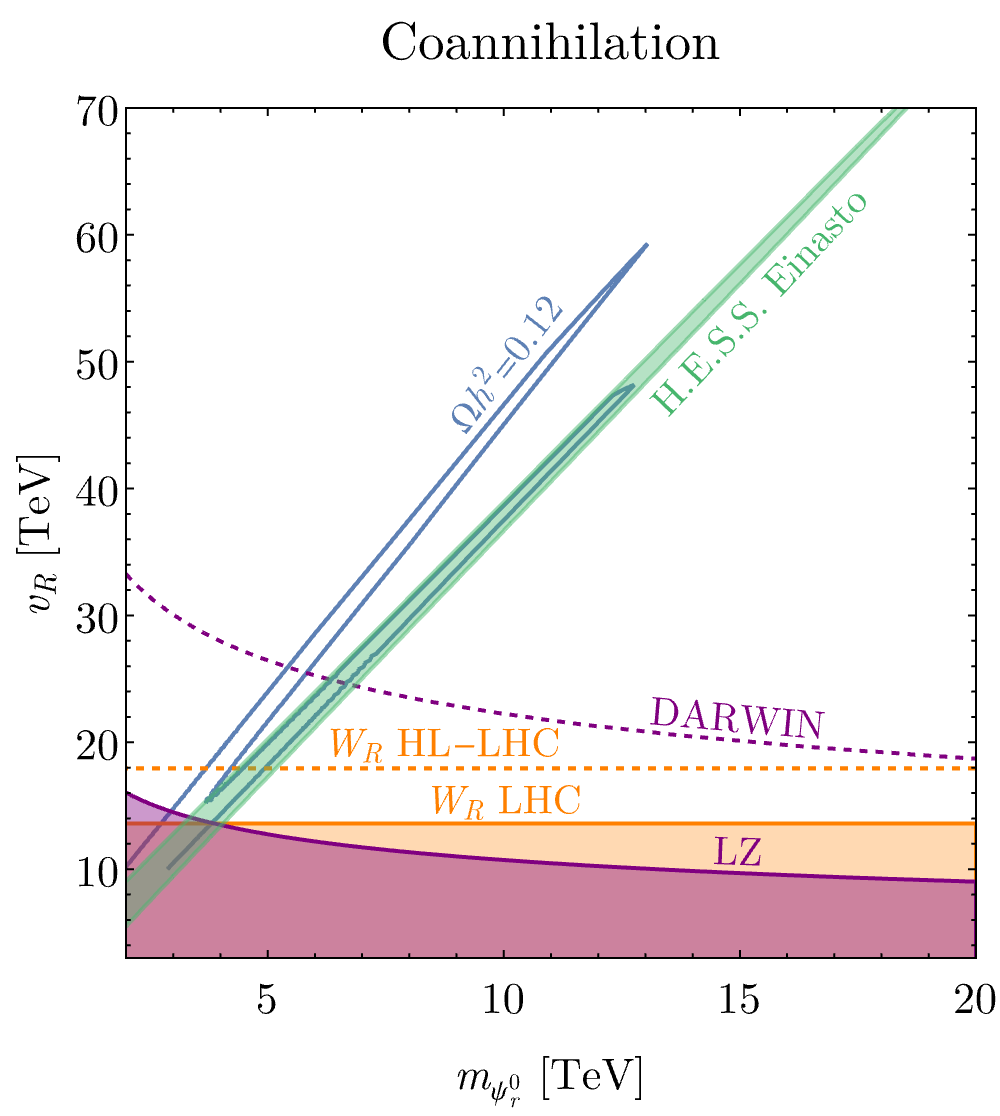}
    \hspace*{0.2cm}
    \includegraphics[width=0.49\linewidth]{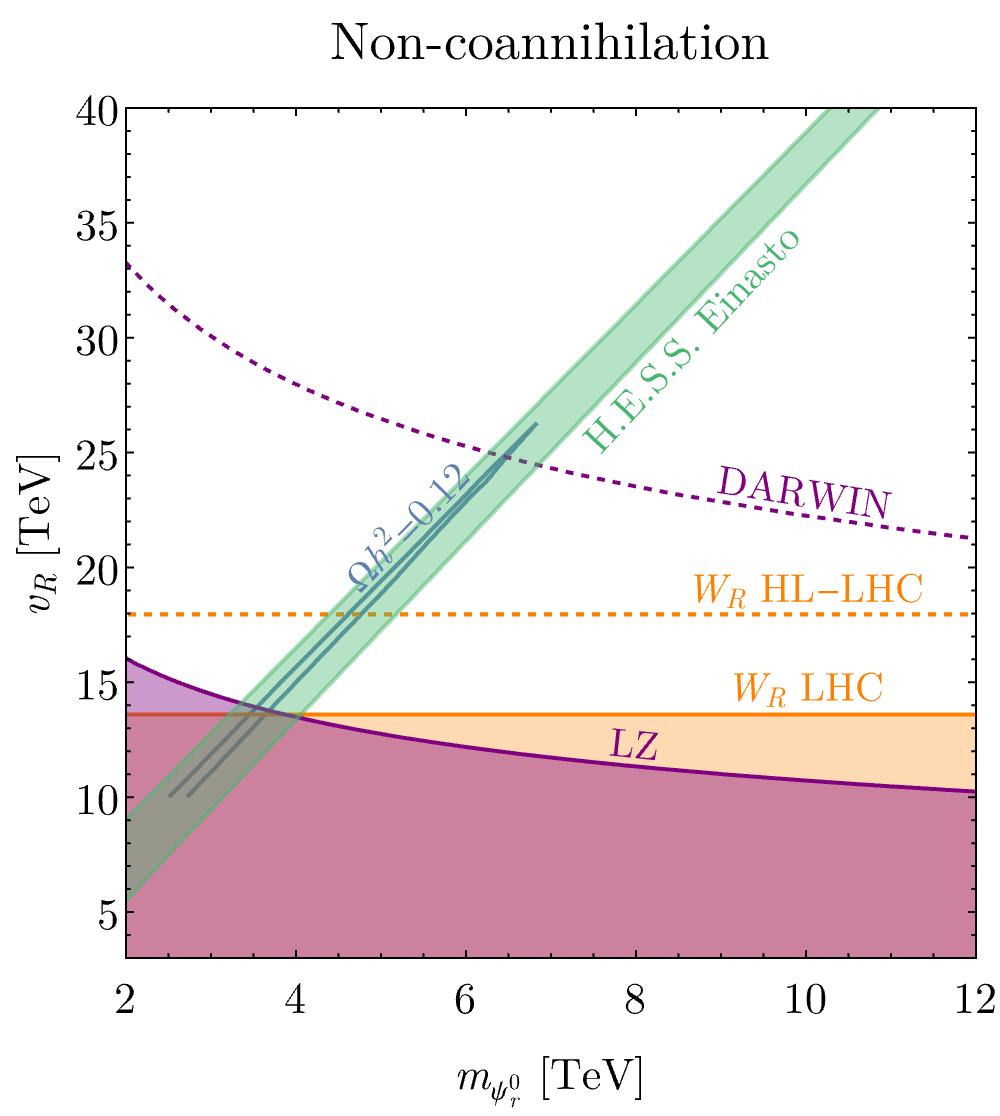}
    \caption{Constraints  on the DM mass $m_{\psi_r^0}$ and the Parity symmetry breaking scale $v_R$.
        Left: coannihilation case. Right: non-coannihilation case. On the blue curves,  the observed DM density is obtained. The shaded regions are excluded by the respectively labeled searches. The dashed curves are future projections. The $Z'$ boson resonance region is ruled out by the H.E.S.S.~experiment assuming the Einasto profile. However, if a more cored profile is assumed, the constraint is relaxed and no parameter space is excluded by indirect-detection experiments.}
    \label{fig:final-pair-doublet}
\end{figure}

Fig.~\ref{fig:doublet-pair-fixvR} shows the DM relic density as a function of the DM mass $m_{\psi_r^0}$ for a fixed $v_R = 15~\tev$.
For the coannihilation case, we observe two resonance peaks, corresponding to $W_R$ and $Z'$ gauge bosons, respectively.
In addition, $m_{\psi_r^0} \simeq 260~\gev$ can also produce the observed relic abundance successfully.
This mass, however, is excluded by searching for $\psi_r^-$ as charged tracks at the LHC~\cite{ATLAS:2019gqq,ATLAS:2025fdm}.
For the non-coannihilation case, only one dip in the total DM relic abundance is observed; $\psi_r$ couples to both $W_R$ and $Z'$, but $\psi_\ell$ does not, and therefore cannot have a resonant annihilation around $m_{\psi_r^0} \simeq m_{W_R}/2$, leading to DM overproduction.
As a result, only the annihilation via $Z'$ resonance can produce the correct DM relic abundance.

Fig.~\ref{fig:final-pair-doublet} shows the relic density curve in the $(m_{\psi_r^0},~v_R)$ plane for both cases.
For the coannihilation case, resonance via both $W_R$ and $Z'$ bosons can produce the correct DM relic abundance.
For the non-coannihilation case, only the $Z'$ resonant annihilation can produce the correct DM relic abundance, as explained above.
Consequently, the left panel shows two DM relic curve branches, while the right panel just has one.

In addition, $\Lambda$ is constrained by big-bang nucleosynthesis for the non-coannihilation case.
Using the constraints derived in~\cite{Kawasaki:2017bqm}, we find that the lifetime of $\psi_\ell^0$ should be shorter than 100 seconds. We obtain
\begin{equation}
    \Lambda <  \begin{cases}
        10^{18}~\mathrm{GeV} \left( \frac{m_{\psi_\ell} - m_{\psi_r^0}}{200~\mathrm{GeV}} \right)^{1/2} \left( \frac{v_R}{20 \mathrm{TeV}} \right) & m_{\psi_\ell} - m_{\psi_r} > m_Z,  \\
        10^{16}~\mathrm{GeV} \left( \frac{m_{\psi_\ell} - m_{\psi_r^0}}{50~\mathrm{GeV}} \right)^{3/2} \left( \frac{v_R}{20 \mathrm{TeV}} \right)  & m_{\psi_\ell} - m_{\psi_r} < m_Z .
    \end{cases}
\end{equation}
One can see that even a cutoff scale as high as $10^{16-18}$ GeV satisfies the constraint.

\section{Experimental signals}
\label{sec:signal}

This section discusses the experimental probes of the DM model.

\subsection{Direct detection}
\label{sec:direct}
The DM $\psi_r^0$ can scatter with nuclei via $Z'$ boson exchange and $Z$ boson exchange with $Z Z'$ mixing, leading to direct detection signals.
The gauge boson exchanges generate the following vector current effective operator between DM, and the up and down quarks,
\begin{align}
    \mathcal{L}_{DD} = &  \frac{1}{6}\sqrt{2} G_{F_R} \left[ \bar{\psi}_r^0 \gamma^\mu \psi_r^0 \bar{u} \gamma_\mu u \left(3-2 t_L^2-8s_L^2 t_L^2\right)  \right.\left. + \bar{\psi}_r^0 \gamma^\mu \psi_r^0 \bar{d}\gamma_\mu d \left(-2t_L^2-3+4s_L^2t_L^2\right) \right],
\end{align}
where $G_{F_R} \equiv 1/(2 \sqrt{2}v_R^2)$ is the $SU(2)_R$ analog of Fermi's constant.
For each quark flavor, the first term corresponds to $Z'$ exchange, and the second term corresponds to $Z Z'$ mixing.
The direct-detection cross section from this operator is
\begin{align}
    \sigma_n =  \frac{G_{F_R}^2 m_n^2}{2 \pi} \frac{1}{A^2} &\left[ (A-Z) \left( 1 + 2t_L^2 \right) \right.\left. - Z \left( 1 - 2 t_L^2 -4s_L^2 t_L^2\right)\right]^2,
\end{align}
where $A$ and $Z$ are the mass number and atomic number of the nucleus, respectively.

The LZ experiment~\cite{LZCollaboration:2024lux} places the most up-to-date constraints on the direct detection signal, corresponding to the purple shaded regions in Fig.~\ref{fig:final-pair-doublet}.
The projected sensitivity of DARWIN~\cite{DARWIN:2016hyl} with an exposure of 1000 ton$\cdot$year is shown by the purple dotted lines in Fig.~\ref{fig:final-pair-doublet}.
For the non-coannihilation case, almost all of the viable parameter space can be probed by DARWIN.

The direct detection signals may be absent if DM becomes a Majorana fermion. In Sec.~\ref{sec:relic pair doublet}, we introduced a higher-dimensional term $\frac{1}{\Lambda}\psi_\ell \psi_r H_L H_R^*$ to allow the $SU(2)_L$ doublet states to decay to DM. If there also exists an operator $\frac{1}{\Lambda'}(\psi_r H_R*)^2$ or $\frac{1}{\Lambda'}(\bar{\psi}_r  H_R)^2$, then $\psi_r^0$ splits into two Majorana states with a mass splitting
\begin{equation}
    \Delta m \simeq \frac{v_R^2}{\Lambda'} = 4~{\rm MeV} \left(\frac{v_R}{20~{\rm TeV}}\right)^2 \frac{10^{11}~{\rm GeV}}{\Lambda'}.
\end{equation}
If $\Lambda' \sim \Lambda$, the coannihilation case leads to $\Delta m \gg 100$ keV and the direct detection signals are absent. In the non-coannihilation case, even if $\Lambda' \sim \Lambda$, we can have $\Delta m $ below $100$ keV.

\subsection{Indirect detection}

DM in the Milky Way can annihilate, leading to indirect detection signals. Since only $\psi_r^0$ exists at the galactic center, the only possible annihilation channel is $\bar{\psi}_r^0 \psi_r^0$ $\to$ SM particles.
The annihilation cross section is computed as
\begin{align}
    \langle \sigma v \rangle_{\bar{\psi}_r^0 \psi_r^0} = \int_0^{v_{\rm esc}} dv (\sigma v)_{\bar{\psi}_r^0 \psi_r^0} 4 \pi v^2 e^{-\frac{v^2}{4v_0^2}}\,,
\end{align}
where $v_0 \simeq 220~\rm km/s$ is the average velocity and  $v_{\rm esc} \simeq 550~\rm km/s$ is the escape velocity. Since only $\bar{\psi}_r^0 \psi_r^0$ annihilation occurs, the resonance is exclusively via $s$-channel $Z'$ boson exchange, and therefore the relic density curve from $W_R$ resonant annihilation cannot be probed via indirect detection.

\begin{figure}[!t]
    \centering
    \includegraphics[width=0.7\linewidth]{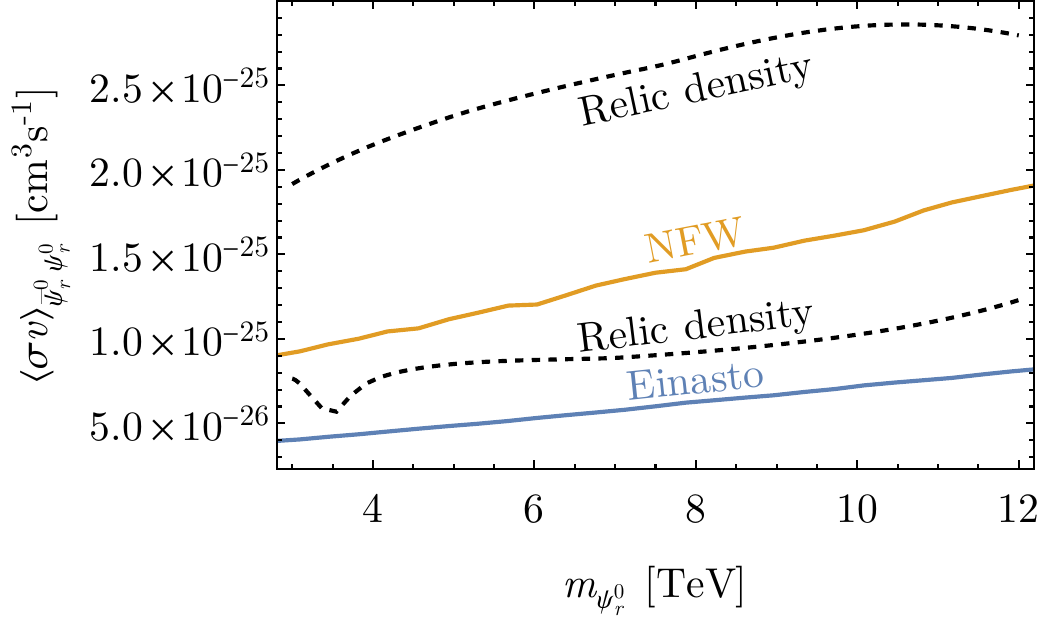}
    \caption{Dark matter annihilation cross section in the Milky Way Galaxy and the upper limits assuming the NFW or Einasto profile from Ref.~\cite{HESS:2022ygk} as a function of the DM mass.
        $v_R$ is chosen so that the DM relic density is correctly produced.
        We choose the coannihilation scenario, but the non-coannihilation case gives a similar annihilation cross section. Two black-dashed lines correspond to two sides of the $Z'$ resonance peak.}
    \label{fig:indi}
\end{figure}

Galactic center gamma ray observations of the Milky Way Galaxy from H.E.S.S.~\cite{HESS:2022ygk} provide the strongest constraints on this indirect detection signal. The green shaded region in Fig.~\ref{fig:final-pair-doublet} is excluded by H.E.S.S. assuming the Einasto profile.
One can see that the $Z'$ boson resonance branch is excluded.
This, however, strongly depends on the DM distribution profile at the galactic center.
If a more cored profile is assumed, then this branch is not excluded.

Fig.~\ref{fig:indi} shows the DM annihilation cross section in the current Milky Way Galaxy and the upper limit assuming Einasto and NFW profiles from H.E.S.S. as a function of the DM mass $m_{\psi_r^0}$, respectively, to show how close the galaxy center annihilation cross section is to the current limit.
The value of $v_R$ is chosen such that the correct DM relic abundance is obtained along the $Z'$ resonance branch.
The coannihilation case is chosen here for its wider range of the DM mass, but the non-coannihilation case predicts a similar cross section.
Two branches correspond to the two sides of the resonance peak of the annihilation.
One can see that both branches are excluded for the Einasto profile, while only one is excluded for the NFW profile.
The predicted annihilation cross section is larger than the upper limit only by an $O(1)$ factor.
Once a more cored profile is assumed, which leads to a much smaller $J$ factor, this $Z'$-resonant annihilation branch completely escapes the search.

On the other hand, if the DM-$Z'$ interaction is inelastic through a Majorana mass term, e.g., the one discussed in Sec.~\ref{sec:direct}, large enough abundances of both the heavier and the lighter states are required for DM annihilation to produce indirect detection signals.
The abundance of the heavier state, however, is suppressed in the present universe, and the indirect detection signal is absent.

\subsection{Collider search}

The predicted $W_R$ and $Z'$ gauge bosons are searched for at the LHC.
A lower bound on $v_R$ can be derived from the gauge boson masses in Eq.~\eqref{eq:gauge boson mass}. Searches for the $W_R$ boson provide the most stringent bound, which exclude $W_R$ bosons with a mass below $6.0$ TeV~\cite{ATLAS:2019lsy}, corresponding to $v_R \simeq 14$ TeV.
The HL-LHC will be able to probe $W_R$ bosons with a mass up to $7.9~\tev$~\cite{ATL-PHYS-PUB-2018-044}, corresponding to $v_R \simeq 18$ TeV. Searches for heavy $Z'$ bosons can be improved by future lepton colliders. For example, the muon collider~\cite{Korshynska:2024suh} is expected to be sensitive to $Z'$ bosons with a mass up to 14 (35) TeV for a 3 (10) TeV muon collider, corresponding to $v_R\simeq 27$ (67) TeV.
The lower bounds on $v_R$ from collider searches are complementary to the upper bound on $v_R$, and future collider experiments are expected to get closer to or even reach the upper bound from the relic abundance.
Even if the direct detection search is evaded, e.g., in the way discussed in Sec.~\ref{sec:direct}, the collider searches provide an unavoidable lower bound on $v_R$.

In addition to the gauge bosons, the DM model predicts new fermions.
$\psi_r^-$ is a long-lived EM charged particle and can be searched for as charged tracks in the LHC.
As stated before, the current bound has excluded the branch with the DM mass around $260~\gev$~\cite{ATLAS:2019gqq}.
The production of $\psi_r^-$ in the resonance region is too small to be observed, even with the production from $\psi_\ell$ decay via $SU(2)_L$ gauge interactions. However, $\psi_r^-$ may be produced from $Z'$ and $W_R$ resonantly. 
For example, with the production cross section of $W_R$ calculated in Ref.~\cite{ATL-PHYS-PUB-2018-044}, one expects more than few of $\psi_r^-$ to be produced at the HL-LHC via $W_R$ resonance for $m_{\psi_r^0}< 5$ TeV  along the  $W_R$ resonance branch.
These particles can then be searched for as charged tracks. The production rate by $Z'$, on the other hand, is smaller by orders of magnitude and cannot be used for $\psi_r^-$ searches.

\section{Summary and conclusions}
\label{sec:summary}

In this paper, we explored doublet WIMP DM in a Parity symmetric theory with minimal Higgs content. The strong CP problem can be solved by Parity. In particular, the model with minimal Higgs content has no physical phases in the field values of the Higgses and can solve the strong CP problem without introducing extra symmetry. The model also does not have an extra hierarchy problem beyond the electroweak hierarchy problem. Doublet WIMP DM may be embedded into the $({\bf 2},{\bf 1},-1/2)\oplus({\bf 1},{\bf 2},-1/2)$ of $SU(2)_L\times SU(2)_R\times U(1)_X$.
From the DM thermal relic abundance, an upper bound on the Parity and $SU(2)_R$ symmetry breaking scale is obtained, as indicated by the abundance predictions in Fig.~\ref{fig:final-pair-doublet}.

Lower bounds from direct-detection and collider experiments complement the upper bound.
$Z'$ exchange contributes to direct detection, which already gives a constraint comparable to the collider constraints.
Future direct-detection experiments will probe the parameter space further. In particular, almost all of the parameter space of the non-coannihilation case can be probed.
On the other hand, current and future collider probes for new gauge bosons provide additional lower bounds on $v_R$, serving as a complementary or alternative approach to direct detection.

An upper bound on the Parity and $SU(2)_R$ breaking scale from WIMP DM phenomenology is encouraging for new-physics searches; without it, the scale may be anywhere between $10$ TeV and $10^{10}$ TeV, and the chance that the Parity and $SU(2)_R$ breaking scale is within the reach of near-future experiments seems low.
Some of the viable parameter spaces have the scale as high as 60 TeV and are not accessible by near-future collider experiments.
However, rare processes such as $\mu\rightarrow e \gamma$ and the electron electric dipole moment arising from the fermion sector may probe such a high symmetry breaking scale, which we leave for future work.

\section*{Acknowledgments}

We thank Joshua Isaacson for useful discussions.
I.\,R.\,W. was supported by Fermi Research Alliance, LLC under Contract No.~DE-AC02-07CH11359 with the U.S. Department of Energy, Office of Science, Office of High Energy Physics, and is currently supported by Fermi Forward Discovery Group, LLC under Contract No. 89243024CSC000002 with the U.S. Department of Energy, Office of Science, Office of High Energy Physics.
I.\,R.\,W. is also supported by DOE distinguished scientist fellowship grant FNAL 22-33.
I.\,R.\,W. is grateful for the support of the New High Energy Theory Center at Rutgers University during the early stage of this project.

\appendix

\section{WIMP embedding in an $SU(2)_L\times SU(2)_R$ bi-doublet}
\label{app:bidoublet}

WIMP DM can be embedded into an $SU(2)_L\times SU(2)_R$ bi-doublet $\psi$ in the $({\bf 2},{\bf 2},0)$ representation of $SU(2)_L\times SU(2)_R\times U(1)_X$. However, as we discuss in this section,
the setup is excluded unless additional particle content is introduced.

The decomposition of $\psi$ under $SU(2)_L\times SU(2)_R$ is
\begin{equation}
    \psi = \begin{pmatrix}
        \psi^0 & \psi^+              \\
        \psi^- & - \overline{\psi}^0
    \end{pmatrix},
\end{equation}
where $\psi^0$ and $\overline{\psi}^0$ are EM neutral, while $\psi^-$ and $\psi^+$ carry EM charges $-1$ and $1$, respectively.
At tree level, their mass terms are given by
\begin{equation}
    {\cal L} =  -\frac{1}{2} m_\psi{\rm tr} \left(i\sigma^2 \psi i \sigma^2 \psi^T \right) + {\rm h.c.} =  -m_\psi \psi^0 \overline{\psi}^0 - m_\psi \psi^- {\psi^+} + {\rm h.c.}.
\end{equation}
$\psi^-$ and $\psi^+$ form a Dirac fermion that we call $\Psi^-$.
At tree level, $\psi_0$ and $\overline{\psi}^0$ also form a Dirac fermion, which we call $\Psi^0$.
In the computation of the DM thermal relic abundance, we may consider both as Dirac fermions.
$\Psi^-$ and $\Psi^0$ have the same mass at tree level.
The mass difference between $\Psi^-$ and $\Psi^0$ arises at the one-loop level after EWSB~\cite{Cirelli:2005uq,Ibe:2012sx},
\begin{align}
    m_{\Psi^-} - m_{\Psi^0} & = \frac{\alpha _2 m_\psi s_L^2 }{4 \pi }f\left(\frac{m_{Z}^2}{m_\psi^2}\right) ,
\end{align}
where $f$ is defined in Eq.~\eqref{eq:loop_f}.
The mass splitting between $\Psi^0$ and $\Psi^-$ is small, and they interact via unsuppressed $W$ boson exchange.
Therefore, coannihilation always occurs.

\begin{figure}[!t]
    \centering
    \includegraphics[width=0.6\linewidth]{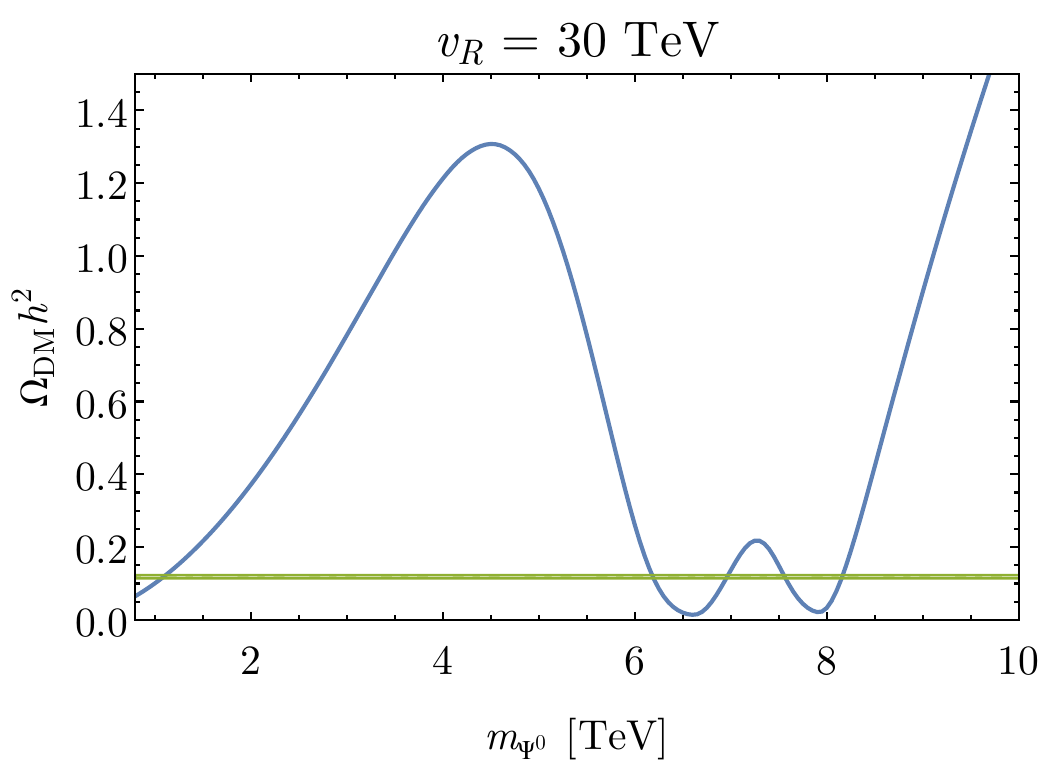}
    \caption{DM relic density for the bi-doublet scenario with $v_R = 30$ TeV. One can see that $m_{\Psi^0} = 1.1$ TeV produces the correct DM abundance, while four additional masses are predicted corresponding to two resonant annihilation branches. The green bands corresponds to $\Omega_{\rm DM}h^2 = 0.1186\pm0.0022$.}
    \label{fig:bidoublet-fixvR}
\end{figure}

\begin{figure}[!t]
    \centering
    \includegraphics[width=0.6\linewidth]{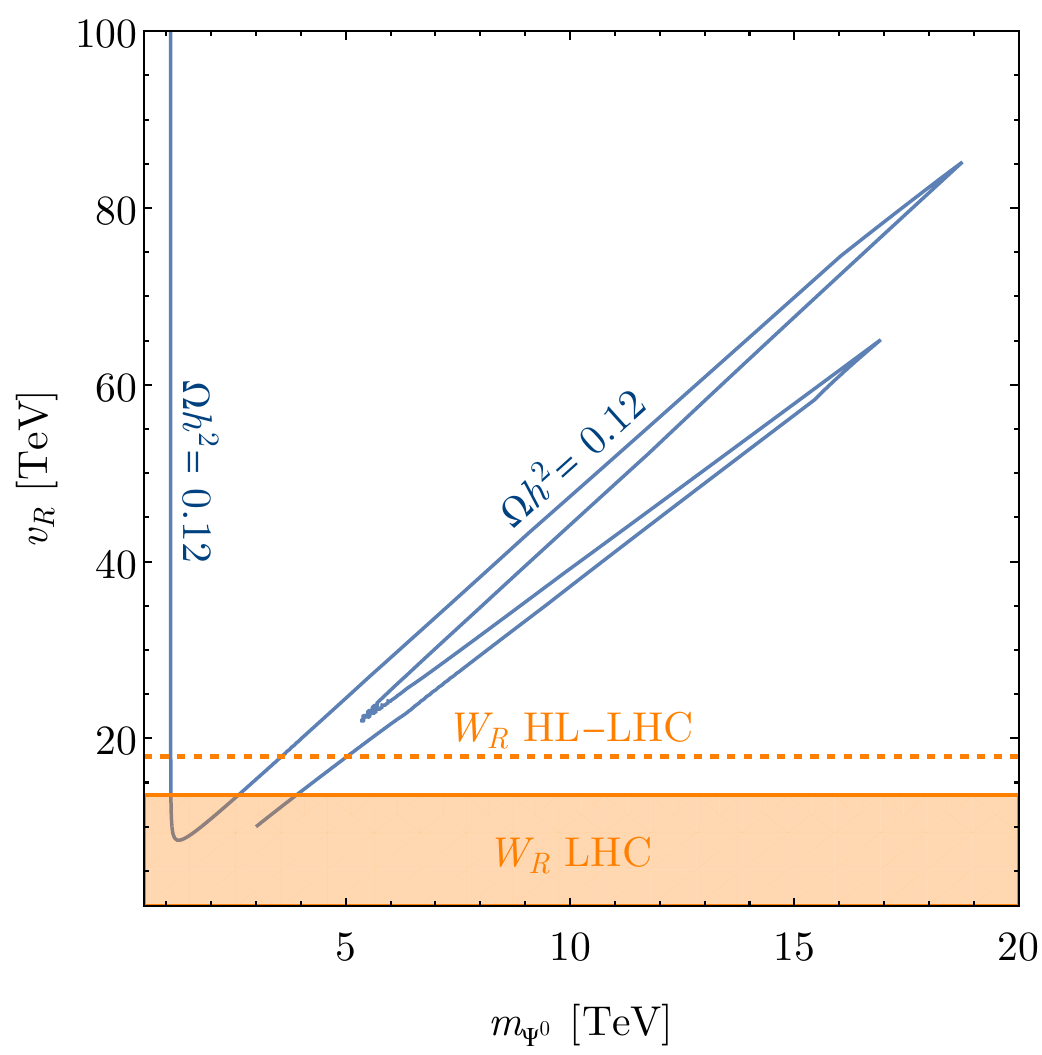}
    \caption{
        Constraints on the  DM mass $m_{\Psi^0}$ and the Parity symmetry breaking scale $v_R$
        for the bi-doublet DM model.
    }
    \label{fig:final bidoublet}
\end{figure}

Fig.~\ref{fig:bidoublet-fixvR} shows the DM relic density as a function of $m_{\Psi^0}$ for fixed $v_R = 30$ TeV.
We observe two resonance peaks: one corresponds to $m_{\Psi^0} = m_{W_R}/2$ and the other one for $m_{\Psi^0} = m_{Z'}/2$.
In addition, $m_{\Psi^0} \simeq 1.1$ TeV, for which $W_R$ and $Z'$ exchange is negligible, is predicted to produce the correct DM relic abundance.
The predicted mass $1.1~\tev$ matches the known result for Higgsino DM production.

Fig.~\ref{fig:final bidoublet} shows the relic density curve for $(\bm{2}, \bm{2}, 0)$ DM in the $(m_{\Psi^0},~v_R)$ plane.
The Parity symmetric theory provides additional viable parameter space through resonant annihilation via $W_R$ and $Z'$ exchange beyond the standard Higgsino scenario.

If $\psi^0$ and $\overline{\psi}^0$ continue to form a Dirac fermion in the late universe, the vector current of DM couples to the $Z$ boson, allowing DM to scatter with nucleons via unsuppressed $Z$-exchange.
Such DM candidates are excluded by direct detection experiments~\cite{Goodman:1984dc,Ahlen:1987mn,Caldwell:1988su,LZCollaboration:2024lux}.
To evade this constraint, the literature introduces mixing between DM and a Majorana fermion to give a Majorana mass term $(\psi^0)^2$ or $(\overline{\psi}^0)^2$, which splits the neutral components of the DM multiplet into two Majorana fermions with a mass splitting $ > O(100)$ keV.
The resulting DM-nucleon scattering cross section is suppressed.
In our set up, tree-level Majorana masses can be generated, for example, by introducing fermions in $({\bf 2},{\bf 1},1/2)$, $({\bf 2},{\bf 1},-1/2)$, $({\bf 1},{\bf 2},1/2)$, $({\bf 1},{\bf 2},-1/2)$, and $({\bf 1},{\bf 1},0)$. These fermions form Yukawa interactions with $\psi$, $H_L$, and $H_R$, and the $({\bf 1},{\bf 1},0)$ fermion may have a Majorana mass term.

\begin{figure}[!t]
    \centering
    \includegraphics[width=0.49\linewidth]{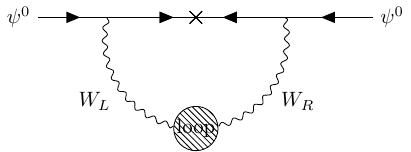}
    \includegraphics[width=0.49\linewidth]{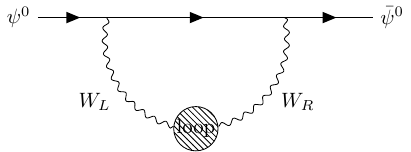}
    \caption{Loop diagrams generating a DM Majorana mass term (left) and kinetic mixing (right).}
    \label{fig:Majorana}
\end{figure}

Alternatively, DM may become a Majorana fermion by quantum corrections.
In fact,
Majorana masses $(\psi^0)^2$ and $(\overline{\psi^0})^2$, and kinetic mixing between $\psi^0$ and $\overline{\psi}^0$ can be generated radiatively as shown in Fig.~\ref{fig:Majorana}. The $W_L$--$W_R$ mixing can arise after $SU(2)_R$ and $SU(2)_L$ breaking.
Ref.~\cite{Garcia-Cely:2015quu} investigates a model where the SM Higgs is embedded into the $({\bf 2},{\bf 2},0)$ of $SU(2)_L\times SU(2)_R\times U(1)_X$, for which the $W_L$--$W_R$ mixing arises at tree level. For the model with $H_L$ and $H_R$ considered in this paper,
the mixing is absent at tree level but can be generated by loop corrections involving $H_L$ and $H_R$.

The diagrams in Fig.~\ref{fig:Majorana} modify the mass and kinetic terms to be
\begin{align}
    \label{eq:mass matrix of majorana}
    \begin{pmatrix}
        \psi^0 & \overline{\psi}^0
    \end{pmatrix}^\dag
    \begin{pmatrix}
        1        & \delta_K \\
        \delta_K & 1
    \end{pmatrix}
    i \bar{\sigma}^\mu \partial_\mu
    \begin{pmatrix}
        \psi^0 \\ \overline{\psi}^0
    \end{pmatrix}+
    \left(
    - \frac{1}{2}
    \begin{pmatrix}
        \psi^0 & \overline{\psi}^0
    \end{pmatrix}
    \begin{pmatrix}
        \delta_M m_\psi & m_\psi          \\
        m_\psi          & \delta_M m_\psi
    \end{pmatrix}
    \begin{pmatrix}
        \psi^0 \\ \overline{\psi}^0
    \end{pmatrix}
    +{\rm h.c.}
    \right)
    ,
\end{align}
where $\delta_K$ and $\delta_M$ should be evaluated on shell.
The mass splitting between the two Majorana fermions is given by
\begin{align}
    \delta m = 2 (\delta_M + \delta_K ) m_\psi.
\end{align}

Let us consider a case where the $W_L$--$W_R$ mixing arises from the sector responsible for the Yukawa interactions.
As an example, we consider an $SO(10)$-inspired setup where the top and bottom Yukawas arise from the following interactions and mass terms,
\begin{align}
    \label{eq:Q Lagrangian}
    {\cal L} = & - x_Q q \bar{Q} H_R - x_Q \bar{q} Q H_L - m_Q Q\bar{Q} + {\rm h.c.}\nonumber \\
               & -x_U q \bar{U} H_L-x_U \bar{q}UH_R-m_U U\bar{U}+ {\rm h.c.}\nonumber         \\
               & -x_D q \bar{D} H_L^*-x_D \bar{q}DH_R^*-m_D D\bar{D}+ {\rm h.c.},
\end{align}
where the gauge charges of $Q$, $\bar{Q}$, $U$, $\bar{U}$, $D$ and $\bar{D}$ are shown in Table~\ref{tab:QUD gauge charges}.
$Q$, $\bar{Q}$, $U$, and $\bar{U}$ can be embedded into the ${\bf 45}$ of $SO(10)$, and $D$ and $\bar{D}$ can be embedded into ${\bf 10}$. To generate the top Yukawa, only one of $(Q,\bar{Q})$ or $(U,\bar{U})$ is required. However, both are necessary in the model in order to generate a Majorana mass splitting that is not suppressed by small $m_Q$, as discussed below.
The top and bottom Yukawas are given by
\begin{equation}
    y_t = \frac{(x_Q^2m_U+x_U^2m_Q)v_R}{\sqrt{m_U^2 + x_U^2 v_R^2 }\sqrt{m_Q^2 + x_Q^2 v_R^2 }},~y_b= \frac{x_D^2 v_R}{\sqrt{m_D^2 + x_D^2 v_R^2}} \frac{m_Q}{\sqrt{m_Q^2 + x_Q^2 v_R^2}}.
\end{equation}
%&

\begin{table}[tbp]
    \caption{The gauge charges of particles in the sector responsible for the Yukawa interactions}
    \begin{center}
        \begin{tabular}{|c|c|c|c|c|c|c|} \hline
                      & $Q$            & $\bar{Q}$        & $U$           & $\bar{U}$       & $D$             & $\bar{D}$       \\ \hline
            $SU(3)_c$ & {\bf 3}        & ${\bf  \bar{3}}$ & {\bf 3}       & ${\bf \bar{3}}$ & {\bf 3}         & ${\bf \bar{3}}$ \\
            $SU(2)_L$ & {\bf 2}        & {\bf 2}          & {\bf 1}       & {\bf 1}         & {\bf 1}         & {\bf 1}         \\
            $SU(2)_R$ & {\bf 2}        & {\bf 2}          & {\bf 1}       & {\bf 1}         & {\bf 1}         & {\bf 1}         \\
            $U(1)_X$  & $-\frac{1}{3}$ & $\frac{1}{3}$    & $\frac{2}{3}$ & $-\frac{2}{3}$  & $- \frac{1}{3}$ & $\frac{1}{3}$   \\ \hline
        \end{tabular}
    \end{center}
    \label{tab:QUD gauge charges}
\end{table}%

\begin{figure}[!t]
    \centering
    \begin{tikzpicture}[baseline=(current bounding box.center)]
        \begin{feynman}
            \vertex (x);
            \vertex[right=0.2\textwidth of x] (z);
            \vertex[right=0.2\textwidth of z] (y);
            \vertex[left=0.1\textwidth of x] (a);
            \vertex[right=0.1\textwidth of y] (d);
            \vertex[above=0.2\textwidth of z] (w);
            \vertex[left=0.1\textwidth of w] (w1);
            \vertex[right=0.1\textwidth of w] (w2);
            \vertex[above=0.1\textwidth of w] (w3);
            \vertex[below=0.1\textwidth of w] (w4);
            \vertex[above=0.01\textwidth of w4] (n4){\(m_D\)};
            \vertex[below=0.01\textwidth of w3] (n4){\(m_U\)};

            \vertex[left=0.1\textwidth of w1] (n5);
            \vertex[right=0.1\textwidth of w2] (n6);

            \vertex[left=0.038\textwidth of w3] (w5){\(\bar{U}\)};
            \vertex[above=0.03\textwidth of w1] (w6){\(q\)};
            \vertex[left=0.04\textwidth of w4] (w7){\(\bar{D}\)};
            \vertex[below=0.032\textwidth of w1] (w8){\(q\)};

            \vertex[right=0.038\textwidth of w3] (w5){\(U\)};
            \vertex[above=0.03\textwidth of w2] (w6){\(\bar{q}\)};
            \vertex[right=0.04\textwidth of w4] (w7){\(D\)};
            \vertex[below=0.032\textwidth of w2] (w8){\(\bar{q}\)};

            \vertex[left=0.071\textwidth of w] (h1);
            \vertex[above=0.071\textwidth of h1] (h2);
            \vertex[above=0.05\textwidth of h2] (h3);
            \vertex[left=0.01\textwidth of h3] (h4) {\(\langle H_L\rangle\)};

            \vertex[right=0.071\textwidth of w] (i1);
            \vertex[above=0.071\textwidth of i1] (i2);
            \vertex[above=0.05\textwidth of i2] (i3);
            \vertex[right=0.01\textwidth of i3] (i4) {\(\langle H_R\rangle\)};

            \vertex[left=0.071\textwidth of w] (j1);
            \vertex[below=0.071\textwidth of j1] (j2);
            \vertex[below=0.05\textwidth of j2] (j3);
            \vertex[left=0.01\textwidth of j3] (j4) {\(\langle H_L^*\rangle\)};

            \vertex[right=0.071\textwidth of w] (k1);
            \vertex[below=0.071\textwidth of k1] (k2);
            \vertex[below=0.05\textwidth of k2] (k3);
            \vertex[right=0.01\textwidth of k3] (k4) {\(\langle H_R^*\rangle\)};

            \diagram*{
            (w1) --[boson, edge label'=\(W_L\)] (n5),
            (n6) --[boson, edge label'=\(W_R\)] (w2),
            (w1) --[quarter left,insertion=0.99] (w3),
            (w2) --[quarter left] (w4),
            (w3) --[quarter left] (w2),
            (w4) --[quarter left,insertion=0.01] (w1),
            (h2) --[scalar] (h4),
            (i2) --[scalar] (i4),
            (j2) --[scalar] (j4),
            (k2) --[scalar] (k4)
            };
        \end{feynman}
    \end{tikzpicture}
    \caption{Mixing between $W_L$ and $W_R$ induced at one loop via the quark sector.}
    \label{fig:WLWR_mixing_UD}
\end{figure}
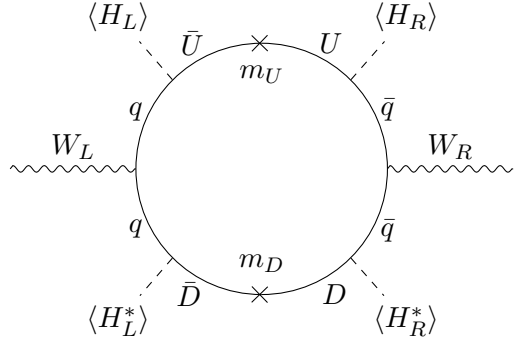

The small bottom Yukawa may be explained by $m_Q \ll v_R$, $m_D \gg v_R$, or $x_D \ll 1$. We choose the first option, for which quantum corrections to the $W_L-W_R$ mixing can be maximized for the following reason.
In the approximation where $\vev{H_L}$ and $\vev{H_R}$ are treated as insertions, $W_L$--$W_R$ mixing that generates the quantum correction to the mass splitting is given in Fig.~\ref{fig:WLWR_mixing_UD}. By taking $x_U\sim x_D = O(1)$, $m_U\sim v_R$, as small as possible $m_D \gtrsim v_R$, and $m_Q \ll v_R$ while satisfying the collider bound on $m_Q$ discussed below, we can maximize the correction. Note that there is another diagram with $Q$ and $\bar{Q}$ inside the loop, but the correction is suppressed by small $m_Q$. The suppression by $m_Q$ can be avoided by taking $m_Q\sim v_R$, but then the small bottom Yukawa needs to be explained by small $x_D$ or large $m_D$, which suppresses the quantum correction.

A lower bound on $m_Q$ is given by collider experiments. The requirement of $x_Q$ being $O(1)$ and $m_Q \ll v_R$ allows us to integrate out heavy states.
The model then predicts an exotic EW doublet charged under the SM gauge group as $(\bm{3}, \bm{2}, -5/6)$, which decomposes into $U'(\bm{3}, 4/3)$ and $D'(\bm{3},-1/3)$ after EWSB.
The masses of these two exotic quarks are $m_Q$, and the $D'$ mixes with the SM bottom quark with the mixing angle
\begin{align}
    \theta_{D'b} \simeq \frac{x_Q v_L}{m_Q}.
\end{align}
Such a mixing leads to the interaction vertex $\bar{U}' \slashed{W} b$.

Searches for $U'$ and $D'$ have been performed in two production channels at the LHC.
A pair production via the strong interaction sets a limit $m_Q > 1.7~\tev$~\cite{ATLAS:2024gyc}.
Single production of $U'$ via $W_Lb$ fusion is also searched for at ATLAS~\cite{ATLAS:2024kgp,ATLAS:2025bzt},
which places $m_Q$-dependent upper bounds on $\theta_{D'b}$. $m_Q > 1.7$ TeV satisfies the bound for $x_Q = 1$.

\begin{figure}[!t]
    \centering
    \includegraphics[width=0.5\linewidth]{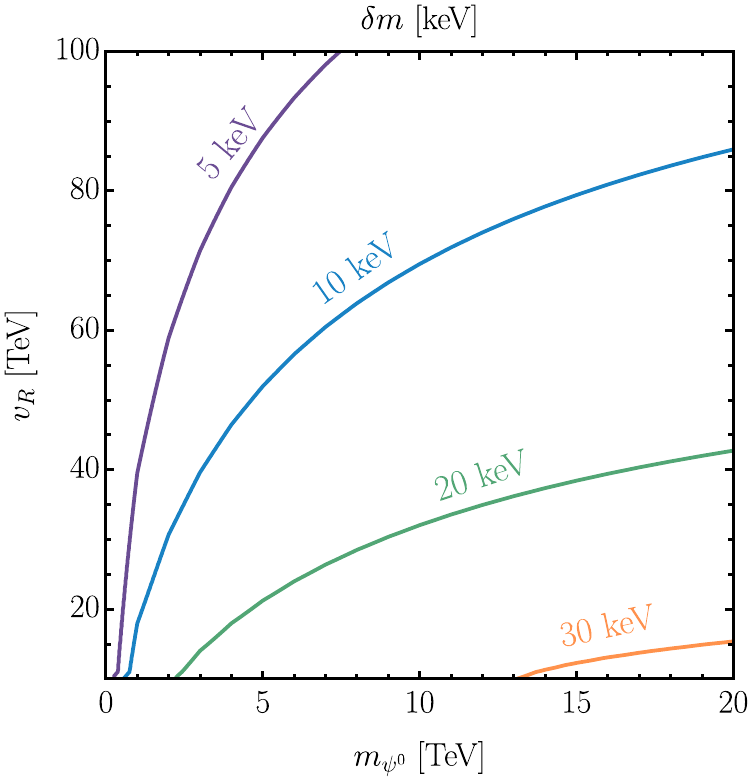}
    \caption{Radiatively generated mass splitting between DM Majorana states for the $SO(10)$-inspired model with $x_Q=x_U=x_D=1$ and $m_Q=2$ TeV. Mass splitting of $O(10)$ keV can be achieved, while an $O(100)$ keV mass splitting is required to avoid direct detection constraints.}
    \label{fig:mMajMassContour}
\end{figure}

\newpage
We perform the computation of the correction without the insertion approximation for $H_R$ but with the insertion approximation for $H_L$. We find
\begin{align}
    \label{eq:mass splitting majorana}
    \delta_M =             & \frac{24 g^2v_L^2 v_R^2 x_U^2 x_D^2 m_U m_D}{(M_U^2-M_Q^2)(M_D^2-M_Q^2)(v_R^2-v_L^2)} \int \frac{d^4k}{(2\pi)^4} \frac{d^4l}{(2\pi)^4} \frac{1}{(p+k)^2 - m_\psi^2} \nonumber                        \\
                           & \times\left(\frac{1}{k^2 - m_{W_R}^2} - \frac{1}{k^2-m_{W_L}^2}\right)
    \left(\frac{1}{l^2 - M_U^2} - \frac{1}{l^2-M_Q^2}\right)\nonumber                                                                                                                                                             \\
                           & \times\left(\frac{1}{(l+k)^2 - M_D^2} - \frac{1}{(l+k)^2-M_Q^2}\right) _{p^2 = m_\psi^2},\nonumber                                                                                                   \\
    \delta_K \slashed{p} = & -\frac{12 g^2v_L^2 v_R^2 x_U^2 x_D^2 m_U m_D}{(M_U^2-M_Q^2)(M_D^2-M_Q^2)(v_R^2-v_L^2)} \int \frac{d^4k}{(2\pi)^4} \frac{d^4l}{(2\pi)^4} \frac{\slashed{p}+\slashed{k}}{(p+k)^2 - m_\psi^2} \nonumber \\
                           & \times\left(\frac{1}{k^2 - m_{W_R}^2} - \frac{1}{k^2-m_{W_L}^2}\right)
    \left(\frac{1}{l^2 - M_U^2} - \frac{1}{l^2-M_Q^2}\right)\nonumber                                                                                                                                                             \\
                           & \times\left(\frac{1}{(l+k)^2 - M_D^2} - \frac{1}{(l+k)^2-M_Q^2}\right) _{p^2 = m_\psi^2},
\end{align}
where $M_i = \sqrt{x_i^2 v_R^2 + m_i^2}$ for $i=Q,U,D$ and we take $m_Q\ll x_Q v_R$ so $M_Q\approx x_Q v_R$.
Here we have performed partial fractions, e.g.,
\begin{equation}
    \frac{1}{(k^2-m_1^2)(k^2-m_2^2)} = \frac{1}{m_1^2-m_2^2}\left(\frac{1}{k^2-m_1^2}- \frac{1}{k^2-m_2^2} \right).
\end{equation}
We introduce Feynman parameters, integrate over $l$ and $k$, and then perform the Feynman integral numerically. The contours of the mass splitting are shown in Fig.~\ref{fig:mMajMassContour}, where we take $x_U=x_Q=x_D =1$ and $m_Q = 2$ TeV.
One can see that only $O(10)$ keV mass splitting can be achieved, which is excluded by direct detection constraints, as discussed in Sec.~\ref{sec:direct}, and therefore the $SO(10)$-inspired model has already been excluded.
Mass splitting of $O(100)$ keV is possible if the Yukawa couplings $x_i$ are as large as $2-3$, but then the coupling becomes non-perturbative just above $v_R$ and theoretical control is lost.

The quantum correction to the mass splitting is small because of the parameter constraints to reproduce the SM Yukawa couplings. By adding extra particles which have nothing to do with the generation of SM Yukawa interactions and couple to $H_L$ and $H_R$, a Majorana mass splitting $> 200$ keV can be obtained, but since the model becomes much more complicated, we do not pursue this possibility further.

\newpage
\section{Gauge interaction}\label{app:gauge_interactions}

In this appendix, we show the gauge interactions of SM fermions, Higgses, and DM multiplets.

\subsection{Mass mixing and the interaction with the SM Higgs}
The SM Higgs $h_L$ couples both to SM gauge bosons and heavy gauge bosons $Z'$.
The coupling to the $W_L$ boson is given by
\begin{align}
    \mathcal{L} \supset g m_{W_L} h_L W_L^+ W_L^- + \frac{g^2}{4} h_L^2 W_L^+ W_L^-.
\end{align}
The coupling to the $Z$ boson is given by
\begin{align}
    \mathcal{L} \supset \frac{1}{8}(g^2 + g_Y^2) h_L^2 Z^2 + g \frac{m_Z^2}{m_{W_L}}h_L Z Z.
\end{align}

Besides these couplings, the $U(1)_X$ charge of $H_L$ also couples it to the $Z'$ boson.
The coupling is
\begin{align}
    \mathcal{L} \supset \frac{1}{2}g\frac{s_R^4}{c_R^2} m_{W_L} h_L Z'^2 + \frac{1}{8} g^2 \frac{s_R^4}{c_R^2} h_L^2 Z'^2.
\end{align}
This $Z'$ can mix with the SM $Z$ boson via the following terms,
\begin{align}
    \mathcal{L} \supset \frac{g^2}{4} \frac{s_L^2}{c_L^2 \sqrt{1-2s_L^2}} h_L^2 Z Z' + g \frac{s_L^2}{c_L \sqrt{1-2s_L^2}}  m_Z h_L Z Z' + \frac{s_L^2}{\sqrt{1-2s_L^2}}  m_Z^2 Z Z'.
\end{align}
The mixing angle $\theta_{Z Z'}$ is generated by the last term,
\begin{align}
    \sin \theta_{Z Z'} \simeq  \frac{m_Z^2}{m_{Z'}^2} \frac{s_L^2}{\sqrt{1-2s_L^2}}.
\end{align}

\subsection{Interaction with SM fermions}

We assume the SM fermions and right-handed neutrinos dominantly come from $q$, $\bar{q}$, $\ell$, and $\bar{\ell}$,
\begin{align}
    q = \begin{pmatrix}
            u \\ d
        \end{pmatrix},~
    \bar{q} = \begin{pmatrix}
                  \bar{d} \\ - \bar{u}
              \end{pmatrix},~
    \ell = \begin{pmatrix}
               \nu \\ e
           \end{pmatrix},~
    \bar{\ell} = \begin{pmatrix}
                     \bar{e} \\ -\bar{N}
                 \end{pmatrix}.
\end{align}
The gauge interactions of SM fermions and right-handed neutrinos are given by
\begin{align}
    \mathcal{L} & =\frac{g}{\sqrt{2}} W_{L\mu}^- \left(d^\dag \bar{\sigma}^\mu u + e^\dag \bar{\sigma}^\mu \nu  \right)+   \frac{g}{\sqrt{2}} W_{L\mu}^+ \left(u^\dag \bar{\sigma}^\mu d + \nu^\dag \bar{\sigma}^\mu e \right) \nonumber                                              \\
                & -   \frac{g}{\sqrt{2}} W_{R\mu}^- \left(\bar{u}^\dag \bar{\sigma}^\mu \bar{d} + \bar{N}^\dag \bar{\sigma}^\mu \bar{e} \right) -\frac{g}{\sqrt{2}} W_{R\mu}^+ \left(\bar{d}^\dag \bar{\sigma}^\mu \bar{u} + \bar{e}^\dag \bar{\sigma}^\mu \bar{N}  \right) \nonumber \\
                & + \frac{g}{c_R}Z'_\mu\sum_f \left(I_{3R,f}-s_R^2 Y_f\right) f^\dag \bar{\sigma}^\mu f  +
    \frac{g}{c_L}Z_\mu\sum_f \left(I_{3L,f}-s_L^2 Q_f\right) f^\dag \bar{\sigma}^\mu f +
    e A_\mu \sum_f Q_f f^\dag\bar{\sigma}^\mu f,
\end{align}
where $f$ are Weyl fermions.
In terms of four-component fields,
\begin{align}
    \mathcal{L} & =\frac{g}{\sqrt{2}}W_{L\mu}^-\left( \bar{d}\gamma^\mu P_L u + \bar{e}\gamma^\mu P_L \nu \right) + \frac{g}{\sqrt{2}}W_{L\mu}^+\left( \bar{u}\gamma^\mu P_L d + \bar{\nu}\gamma^\mu P_L e \right) \nonumber \\
                & +  \frac{g}{\sqrt{2}}W_{R\mu}^-\left( \bar{d}\gamma^\mu P_R u + \bar{e}\gamma^\mu P_R N \right) + \frac{g}{\sqrt{2}}W_{R\mu}^+\left( \bar{u}\gamma^\mu P_R d + \bar{N }\gamma^\mu P_R e \right) \nonumber  \\
                & + \frac{g}{c_R}Z_\mu' \sum_F \bar{F} \gamma^\mu \left(s_R^2\left(I_{3F}- Q_F\right)P_L + \left(I_{3F}-s_R^2Q_F\right) P_R \right) F \nonumber                                                              \\
                & +  \frac{g}{c_L} Z_\mu \sum_F \bar{F} \gamma^\mu \left(\left(I_{3F}- s_L^2 Q_F\right)P_L - s_L^2Q_F P_R \right) F +
    e A_\mu \sum_F Q_F \bar{F}\gamma^\mu F,
\end{align}
where $F$ are Dirac fields.
We assume that the right-handed neutrinos $\bar{N}$ are lighter than DM so that DM can annihilate into it. It is then convenient to put $\nu$ and $\bar{N}$ into a Dirac field even if they do not have a Dirac mass term.

\subsection{Interaction with doublet pair WIMP dark matter}\label{app:gauge_int_psilr}
The DM multiplets in the doublet pair $({\bf 2},{\bf 1},-1/2)\oplus({\bf 1},{\bf 2},-1/2)$ representation of $SU(2)_L\times SU(2)_R\times U(1)_X$, which we call $\psi_\ell\oplus\psi_r$,
are decomposed as
\begin{equation}
    \psi_\ell = \begin{pmatrix}
        \psi_\ell^0 \\
        \psi_\ell^-
    \end{pmatrix},\
    \psi_r = \begin{pmatrix}
        \psi^0_r \\  \psi_r^-
    \end{pmatrix},~~
\end{equation}
(see Sec.~\ref{sec:WIMPindoubletPair}).
Their gauge interactions are given by
\begin{align}
    \mathcal{L}=g_L\overline{\psi}_\ell\gamma^\mu W_{L\mu}\psi_\ell+g_R\overline{\psi}_r\gamma^\mu W_{R\mu}\psi_r-\frac{1}{2}g_XB_{X\mu}(\overline{\psi}_\ell \gamma^\mu\psi_\ell+\overline{\psi}_r \gamma^\mu\psi_r).
\end{align}
In terms of the Dirac fields $\psi_\ell^0$, $\psi^-_\ell$, $\psi^0_r$ and $\psi^-_r$,
\begin{align}
    \mathcal{L} & =\frac{g_L}{\sqrt{2}}W_{L\mu}^+\overline{\psi^0_\ell}\gamma^\mu\psi^-_\ell+\frac{g_L}{\sqrt{2}}W_{L\mu}^- \overline{\psi^-_\ell}\gamma^\mu\psi^0_\ell                                                                                                                               \nonumber   \\
                & +\frac{g_R}{\sqrt{2}}W_{R\mu}^+\overline{\psi^0_r}\gamma^\mu\psi^-_r+\frac{g_R}{\sqrt{2}}W_{R\mu}^- \overline{\psi^-_r}\gamma^\mu\psi^0_r                                                                                                                                             \nonumber  \\
                & +\frac{g_R}{c_R}Z'_\mu\left(\frac{1}{2}s_R^2\overline{\psi_\ell^0}\gamma^\mu\psi_\ell^0 + \frac{1}{2}s_R^2\overline{\psi_\ell^-}\gamma^\mu\psi_\ell^- + \frac{1}{2}\overline{\psi_r^0}\gamma^\mu\psi_r^0 +\frac{1}{2}\left(2s_R^2-1\right)\overline{\psi_r^-}\gamma^\mu\psi_r^-\right)  \nonumber \\
                & +\frac{g_L}{c_L}Z_\mu\left( \frac{1}{2}\overline{\psi_\ell^0}\gamma^\mu\psi_\ell^0 + \frac{1}{2}\left(2s_L^2-1\right)\overline{\psi_\ell^-}\gamma^\mu\psi_\ell^-  +s_L^2\overline{\psi_r^-}\gamma^\mu\psi_r^-\right)                                                                 \nonumber   \\
                & -eA_\mu\left(\overline{\psi_\ell^-}\gamma^\mu\psi_\ell^- +\overline{\psi_r^-}\gamma^\mu\psi_r^-\right),
\end{align}
where
\begin{align}
    c_R=\frac{\sqrt{g_R^2-g_Y^2}}{g_R},~~t_R=\frac{g_Y}{\sqrt{g_R^2-g_Y^2}},~~s_R = c_R t_R = t_L.
\end{align}

\subsection{Interaction with bi-doublet WIMP dark matter}

The DM multiplet in the bi-doublet $({\bf 2},{\bf 2},0)$ representation of $SU(2)_L\times SU(2)_R\times U(1)_X$, which we call $\psi$,
is decomposed as
\begin{equation}
    \psi = \begin{pmatrix}
        \psi^0 & \psi^+              \\
        \psi^- & - \overline{\psi}^0
    \end{pmatrix},
\end{equation}
(see Appendix~\ref{app:bidoublet}).
The gauge interactions of $\psi$ are given by
\begin{align}
    \mathcal{L}={\rm tr}\left(\psi^\dag \bar{\sigma}^\mu \left(g_LW_{L\mu} \psi -g_R \psi W_{R\mu}\right) \right),
\end{align}
where tr is the trace over the gauge indices. In components,
\begin{align}
    \mathcal{L}&=\frac{g_L}{\sqrt{2}}W_{L\mu}^-\left(
    \psi^{-\dag} \bar{\sigma}^\mu \psi^0
    - \overline{\psi}^{0\dag} \bar{\sigma}^\mu \psi^+
    \right) +  \frac{g_L}{\sqrt{2}}W_{L\mu}^+\left(
    \psi^{0\dag} \bar{\sigma}^\mu \psi^-
    - \psi^{+\dag} \bar{\sigma}^\mu \overline{\psi}^0
    \right) \nonumber  \\
    &+\frac{g_R}{\sqrt{2}}W_{R\mu}^-\left(
    \psi^{-\dag} \bar{\sigma}^\mu \overline{\psi}^0
    - \psi^{0\dag} \bar{\sigma}^\mu \psi^+
    \right) +  \frac{g_R}{\sqrt{2}}W_{R\mu}^+\left(
    \overline{\psi}^{0\dag} \bar{\sigma}^\mu \psi^-
    - \psi^{+\dag} \bar{\sigma}^\mu \psi^0
    \right) \nonumber  \\
    &+\frac{1}{2}g_R c_R Z'_\mu \left(
    - \psi^{0\dag}\bar{\sigma}^\mu \psi^0 + \overline{\psi}^{0\dag}\bar{\sigma}^\mu \overline{\psi}^0 - \psi^{-\dag}\bar{\sigma}^\mu \psi^- + \psi^{+\dag}\bar{\sigma}^\mu \psi^+
    \right)
    \nonumber          \\
    &+ \frac{g_L}{c_L}Z^\mu \left( \frac{1}{2} \psi^{0\dag} \bar{\sigma}^\mu\psi^0 - \frac{1}{2} \overline{\psi}^{0\dag} \bar{\sigma}^\mu\overline{\psi}^0 +
    \left(- \frac{1}{2}+ s_L^2\right ) \psi^{-\dag} \bar{\sigma}^\mu\psi^- +
    \left(\frac{1}{2}- s_L^2\right ) \psi^{+\dag} \bar{\sigma}^\mu\psi^+
    \right)  \nonumber \\
    &+e A_\mu \left(\psi^{+\dag}\bar{\sigma}^\mu \psi^+ -\psi^{-\dag}\bar{\sigma}^\mu \psi^-\right).
\end{align}
We construct the Dirac fields via
\begin{equation}
    \Psi^0 = \begin{pmatrix}
        \psi^0 \\ i \sigma^2 \overline{\psi}^{0*}
    \end{pmatrix},~~
    \Psi^- = \begin{pmatrix}
        \psi^- \\ i \sigma^2 \psi^{+*}
    \end{pmatrix}.
\end{equation}
\newpage
In terms of $\Psi^0$ and $\Psi^-$, the gauge interactions are

\begin{align}
    \mathcal{L} & =\frac{g_L}{\sqrt{2}}W_{L\mu}^-\overline{\Psi}^- \gamma^\mu \Psi^0 + \frac{g_L}{\sqrt{2}}W_{L\mu}^+\overline{\Psi}^0 \gamma^\mu \Psi^-
    \nonumber                                                                                                                                                    \\
                & - \frac{g_R}{\sqrt{2}}W_{R\mu}^-\overline{\Psi}^0 \gamma^\mu C \overline{\Psi}^{-t}+\frac{g_R}{\sqrt{2}}W_{R\mu}^+\Psi^{-t}C \gamma^\mu \Psi^0  \nonumber \\
                & -\frac{1}{2}g_R c_RZ_\mu' \left(
    \overline{\Psi}^0 \gamma^\mu  \Psi^0  +\overline{\Psi}^- \gamma^\mu  \Psi^-
    \right)
    + \frac{g_L}{c_L}Z^\mu \left(\frac{1}{2}\overline{\Psi}^0 \gamma^\mu  \Psi^0
    +  \left(- \frac{1}{2}+ s_L^2\right) \overline{\Psi}^- \gamma^\mu  \Psi^-
    \right)                                                                                                                                                     \nonumber  \\
                & - e A_\mu \overline{\Psi}^- \gamma^\mu \Psi^-.
\end{align}
Defining
\begin{equation}
    \Psi^+\equiv-i\gamma^2\Psi^{-*}=-i\gamma^2\gamma^0 \overline{\Psi}^{-t}=\begin{pmatrix}
        \psi^+ \\ i \sigma^2 \psi^{-*}
    \end{pmatrix}\equiv -C\overline{\Psi}^{-t},
\end{equation}
where $C\equiv i\gamma^2\gamma^0$, the gauge interactions are
\begin{align}
    \mathcal{L} & =\frac{g_L}{\sqrt{2}}W_{L\mu}^-\overline{\Psi}^- \gamma^\mu \Psi^0 + \frac{g_L}{\sqrt{2}}W_{L\mu}^+\overline{\Psi}^0 \gamma^\mu \Psi^-
    \nonumber                                                                                                                                            \\
                & +\frac{g_R}{\sqrt{2}}W_{R\mu}^-\overline{\Psi}^0 \gamma^\mu \Psi^+ +\frac{g_R}{\sqrt{2}}W_{R\mu}^+\overline{\Psi}^+ \gamma^\mu \Psi^0   \nonumber \\
                & -\frac{1}{2}c_R g_RZ_\mu' \left(
    \overline{\Psi}^0 \gamma^\mu  \Psi^0  +\overline{\Psi}^- \gamma^\mu  \Psi^-
    \right)
    + \frac{g_L}{c_L}Z^\mu \left(\frac{1}{2}\overline{\Psi}^0 \gamma^\mu  \Psi^0
    +  \left(- \frac{1}{2}+ s_L^2\right) \overline{\Psi}^- \gamma^\mu  \Psi^-
    \right)                                                                                                                                              \nonumber \\
                & - e A_\mu \overline{\Psi}^- \gamma^\mu \Psi^-.
\end{align}
For convenience, we may also write the interactions with $W_R$ in the following form:
\begin{align}
      \frac{g_R}{\sqrt{2}}W_{R\mu}^-\overline{\Psi}^0 \gamma^\mu \Psi^+ +\frac{g_R}{\sqrt{2}}W_{R\mu}^+\overline{\Psi}^+ \gamma^\mu \Psi^0 & = \frac{g_R}{\sqrt{2}} W_{R\mu}^-\overline{\Psi}^- \gamma^\mu C \overline{\Psi}^{0t}-\frac{g_R}{\sqrt{2}} W_{R\mu}^+ \Psi^{0t}C \gamma^\mu \Psi^-.
\end{align}

\section{One-loop mass difference from gauge interactions}
\label{sec:quantum correction}
In this appendix, we compute the one-loop quantum corrections to the mass difference between DM multiplets from gauge interactions.

The quantum correction to the masses can be calculated as follows.
We first define the fermion self-energy as
\begin{align}
    \Sigma(p) = \slashed{p} \Sigma_K + \Sigma_M.
\end{align}
The fermion propagator with mass $m_\psi$ can then be written as
\begin{align}
    S(p) = \frac{1}{1+ \Sigma_K} \frac{1}{\slashed{p} - \frac{1+\Sigma_M/m_\psi}{1 - \Sigma_K}m_\psi}.
\end{align}
Here, $\Sigma_K$ represents the wave function renormalization.
The pole mass can thus be defined as
\begin{align}
    M_p = \mathrm{Re}\left(\frac{1+\Sigma_M(m_\psi)/m_\psi}{1 - \Sigma_K(m_\psi)}m_\psi\right),
\end{align}
which generates the quantum correction
\begin{align}
    \delta m_\psi = - \mathrm{Re} \left(\Sigma_M(m_\psi) + m_\psi \Sigma_K(m_\psi) \right).
\end{align}

Next we evaluate the self-energy diagrams. We remind the reader of the Passarino-Veltman integrals defined in $D$-dimension with $D \equiv 4 - 2 \epsilon$ under dimensional regularization.
Those associated with the one-loop fermion self-energy are defined as
\begin{align}
    B_0(p^2, m_1, m_\psi) =       & \int d^Dk \frac{1}{[k-m_1]^2[(k+p)^2 - m_\psi^2]},      \\
    p_\mu B_1(p^2, m_1, m_\psi) = & \int d^D k \frac{k^\mu}{[k-m_1]^2[(k+p)^2 - m_\psi^2]}.
\end{align}

\begin{figure}[!t]
    \centering
    \includegraphics[width=0.6\linewidth]{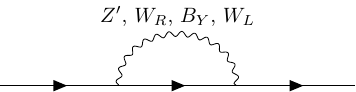}
    \caption{Loop diagram for fermion self-energy. Diagram made with \textsc{Tikz-Feynman}~\cite{Ellis:2016jkw}.}
    \label{fig:1-loop mass}
\end{figure}

We use \verb|Package-X|~\cite{Patel:2015tea,Patel:2016fam} to perform the loop evaluation.
There is only one type of diagram to evaluate, shown in Fig.~\ref{fig:1-loop mass}.
This diagram can be evaluated for general gauge boson mass $m_V$ and coupling $g$ as
\begin{align}
    \Sigma_M & = \frac{g^2}{16 \pi^2}D m_\psi B_0 (p^2, m_\psi, m_V),    \\
    \Sigma_K & = \frac{g^2}{16 \pi^2}(D-2) m_\psi B_1(p^2, m_\psi, m_V).
\end{align}
For massless gauge bosons, we can safely take $m_V = 0$ in the above expressions.
We work in the case where $SU(2)_R \times U(1)_X$ is broken but EW symmetry is unbroken.
In this case, $B_Y$ and $W_L$ are massless while $Z'$ and $W_R$ are massive.
For convenience, we define the quantum correction to the mass as
\begin{align}
    \delta M(m_V) = - \mathrm{Re}\left(\Sigma_M (p^2 = m_\psi^2) + m_\psi \Sigma_K(p^2 = m_\psi^2) \right),
\end{align}
and identify $g$ as $g_L = g_R$ in the following.
We define $\alpha_2 \equiv g^2/(4\pi)$.

Particle $\psi_r^0$ receives quantum corrections from $Z'$ and $W_R$ gauge bosons; $\psi_r^-$ from $Z'$, $W_R$, and $B_Y$; and
$\psi_\ell$ from $Z'$, $B_Y$, and $W_L$. These quantum corrections are
\begin{align}
    \delta m_{\psi_r^0}  & = \frac{\alpha_2}{4 \pi} \left( \frac{1}{2} \delta M(m_{W_R}) + \frac{1}{4 c_R^2} \delta M(m_{Z'})  \right),                                        \\
    \delta m_{\psi_r^-}  & = \frac{\alpha_2}{4 \pi} \left(\frac{1}{2} \delta M(m_{W_R}) + (\frac{1}{c_R} (s_R^2 - \frac{1}{2}))^2 \delta M(m_{Z'}) + s_R^2 \delta M(0) \right) ,\\
    \delta m_{\psi_\ell} & =  \frac{\alpha_2}{4 \pi} \left( \left( \frac{s_R^2}{2 c_R} \right)^2 \delta M(m_{Z'}) + C_F \delta M(0) + \frac{1}{4} s_R^2 \delta M(0) \right).
\end{align}
Here, $C_F = 3/4$ is the Casimir operator of the fundamental representation of $SU(2)$. The mass differences in Eq.~\eqref{eq:mass splitting pair doublet} follow from these quantum corrections.

\section{Annihilation channel of DM multiplets}
\label{sec:channel}

In this appendix, we show the annihilation cross sections of DM multiplets.
We take the non-relativistic limit and compute s-wave annihilation cross sections.
The cross sections and corresponding long-range potentials are shown for different initial spin states, for both the doublet pair $({\bf 2}, {\bf 1},-1/2)\oplus({\bf1}, {\bf 2},-1/2)$ and bi-doublet $({\bf 2}, {\bf 2},0)$ WIMP models.
To simplify the expressions, the cross sections are shown in the limit of vanishing SM particle masses, while the true masses are applied in numerical computation.
The Mandelstam variable $s$ is directly set to $4 m_\psi^2$ for conciseness of expressions.
Furthermore, the Breit-Wigner width is not shown here. To compute the annihilation cross sections, we make use of the FeynCalc~\cite{Shtabovenko:2023idz,Shtabovenko:2020gxv,Shtabovenko:2016sxi,Mertig:1990an}, FeynRules~\cite{Alloul:2013bka} and FeynArts~\cite{Hahn:2000kx} Mathematica packages.

\subsection{Doublet pair WIMP dark matter}

\subsubsection{$\bar{\psi}_r^0 \psi_r^0$ annihilation}
The only non-negligible channels for $ \bar{\psi}_r^0 \psi_r^0$ annihilation are $s$-channel annihilation into $\bar{f}f$ or $Z h_L$ with the initial state having total spin 1.
The cross sections are
\begin{align}
    \sigma v|_{\bar{f}f} & = \frac{\pi \alpha_2^2}{6 c_R^4 m_\psi^2} \left( 1 - \frac{m_{Z'}^2}{4m_\psi^2} \right)^{-2} \left( I_{3f}^2(1+s_R^4) - 2 I_{3f} Q_f(1+s_R^2) s_R^2 + 2 Q_f^2 s_R^4 \right), \nonumber \\
    \sigma v|_{Zh_L}       & = \frac{\pi \alpha_2^2 s_L^4}{48 c_R^2 c_L^2 (2c_L^2 -1) m_\psi^2} \left( 1 - \frac{m_{Z'}^2}{4m_\psi^2} \right)^{-2},
\end{align}
where $Q_f$ and $I_{3f}$ are the EM charge and isospin of fermion $f$, respectively.

There is no non-negligible long-range force in such an initial state, and thus the Sommerfeld effect is negligible.

\subsubsection{$\psi_r^+ \psi_r^-$ annihilation}
When the total spin of the initial state is 0, the annihilation cross sections are
\begin{align}
    \sigma v|_{\gamma \gamma} & = \frac{4 \pi \alpha_2^2 s_L^4}{m_\psi^2},~~\sigma v_{Z \gamma}        = \frac{8 \pi  \alpha _2^2 s_L^6}{m_\psi^2 c_L^2},~~ \sigma v|_{Z Z}            = \frac{4 \pi  \alpha _2^2 s_L^8}{c_L^4 m_\psi^2}, \nonumber   \\
    \sigma v|_{Z' \gamma}     & =\frac{8 \pi \alpha_2^2 s_L^2 (s_R^2-\frac{1}{2})^2}{c_R^2 m_\psi^2} \left( 1 - \frac{m_{Z'}^2}{2m_\psi^2} + \frac{m_{Z'}^4}{16m_\psi^4} \right)^{3/2} \left( 1 - \frac{m_{Z'}^2}{4m_\psi^2} \right)^{-2},  \nonumber \\
    \sigma v|_{Z Z'}          & = \frac{2 \pi  \alpha _2^2 \left(1-2 s_R^2\right)^2 s_L^4 }{c_R^2 c_L^2 m_\psi^2} \left( 1 - \frac{m_{Z'}^2}{4m_\psi^2} \right)^3 \left( 1 - \frac{m_{Z'}^2}{4m_\psi^2} + \frac{m_{Z'}^4}{32 m_\psi^4} \right)^{-2}.
\end{align}

When the total spin of the initial state is 1, the annihilation cross sections are
\begin{align}
    \sigma v|_{W_LW_L}       & = \frac{\pi \alpha_2^2 s_L^4}{12 c_L^4 m_\psi^2} \left[ 1+ \frac{c_L (1-2s_R^2)}{c_R \sqrt{2c_L^2 - 1}} \left( 1 - \frac{m_{Z'}^2}{4m_\psi^2} \right)^{-1} + \frac{c_L^2 (1-2s_R^2)^2}{4 c_R^2 (2c_L^2 - 1)}  \left( 1 - \frac{m_{Z'}^2}{4 m_\psi^2} \right)^{-2}\right] \,,                                                                               \nonumber \\
    \sigma v|_{\bar{f}f} & = \frac{2 \pi \alpha_2^2}{3 m_\psi^2} \left[ \frac{s_L^4}{c_L^4} \left( I_{3f}^2 + 2 I_{3f} Q_f (1- 2s_L^2) + 2 Q_f^2 (1-2s_L^2)^2 \right) \right. \nonumber                                                                                                                                                                                                         \\
                         & \left. - \frac{s_L^2 (1-2s_R^2)}{c_R^2 c_L^2} \left( 1 - \frac{m_{Z'}^2}{4m_\psi^2} \right)^{-1} \left( Q_f (1 - 2 s_L^2) (I_{3f} + s_R^2(I_{3f} - 2 Q_f)) + I_{3f} s_R^2 (I_{3f} - Q_f)\right)\right.\nonumber                                                                                                                                                      \\
                         & \left. + \frac{(1-2s_R^2)^2}{4c_R^4} \left( 1 - \frac{m_{Z'}^2}{4m_\psi^2} \right)^{-2} \left( I_{3f}^2 - 2 I_{3f} Q_f s_R^2 + s_R^4 \left( I_{3f}^2 - 2 I_{3f} Q_f + 2Q_f \right) \right)\right],                                                             \nonumber                                                                                             \\
    \sigma v|_{Zh_L}       & = \frac{\pi \alpha_2^2 s_L^4}{12 c_L^4 m_\psi^2} \left[ 1 - \frac{c_L (1-2s_R^2)}{c_R \sqrt{2c_L^2-1}}\left( 1 - \frac{m_{Z'}^2}{4m_\psi^2} \right)^{-1} + \frac{c_L^2 (1-2s_R^2)^2}{4c_R^2 (2c_L^2-1)} \left(1- \frac{m_{Z'}^2}{4m_\psi^2} \right)^{-2} \right].
\end{align}

The long-range potential of $\psi_r^+ \psi_r^-$ annihilation is mediated by the photon and $Z$ boson,
\begin{align}
    V(r) = - \frac{\alpha}{r} - \frac{\alpha_2}{r} \frac{s_L^4}{c_L^2} e^{- m_Z r},
\end{align}
where $\alpha=e^2/(4\pi)$.

\subsubsection{$\bar{\psi}_r^0 \psi_r^-$ annihilation}
When the total spin of the initial state is 0, the annihilation cross sections are
\begin{align}
    \sigma v|_{W_R \gamma} & = \frac{\pi  \alpha _2^2 s_L^2}{m_\psi^2} \left( 1 - \frac{m_{W_R}^2}{4m_\psi^2}\right),~~
    \sigma v|_{W_R Z}       = \frac{\pi \alpha_2^2 s_L^4}{c_L^2 m_\psi^2} \left( 1 - \frac{m_{W_R}^2}{4m_\psi^2} \right).
\end{align}

When the total spin of the initial state is 1, the annihilation cross sections are
\begin{align}
    \sigma v|_{W_R Z}     & = \frac{\pi \alpha_2^2 s_L^4}{24 c_L^2 m_\psi^2} \left( \frac{m_Z}{m_{W_R}} \right)^2 \left( 1 - \frac{m_{W_R}^2}{4m_\psi^2} \right)^{-1} \left( 1 + \frac{5 m_{W_R}^2}{2 m_\psi^2} + \frac{m_{W_R}^4}{m_\psi^4} \right)\,, \nonumber \\
    \sigma v|_{\bar{f}f'} & = \frac{\pi \alpha_2^2}{6 m_\psi^2} \left(1 - \frac{m_{W_R}^2}{4m_\psi^2}\right)^{-2}.
\end{align}

There is no long-range force mediated by EW gauge bosons for this initial state, and thus the Sommerfeld effect is negligible.

\subsubsection{$\bar{\psi}_l^0 \psi_\ell^0$ and $\psi_\ell^+ \psi_\ell^-$ mixed annihilation}
When the initial state $\psi_\ell^0 \psi_\ell^0$ total spin is 0, the annihilation cross sections are
\begin{align}
    \sigma v |_{W_LW_L}  & = \frac{\pi \alpha_2^2}{2 m_\psi^2},~~
    \sigma v |_{ZZ}   = \frac{\pi \alpha_2^2}{4 c_L^4 m_\psi^2},                                                                                                                                          \nonumber \\
    \sigma v|_{Z Z'} & = \frac{\pi \alpha_2^2 s_R^4}{2 c_R^2 c_L^2 m_\psi^2} \left( 1 - \frac{m_{Z'}^2}{4m_\psi^2} \right)^3 \left( 1 - \frac{m_{Z'}^2}{4m_\psi^2} + \frac{m_{Z'}^4}{32 m_\psi^4} \right)^{-2}.
\end{align}

When the initial state $\psi_\ell^0 \psi_\ell^0$ total spin is 1, the annihilation cross sections are
\begin{align}
    \sigma v |_{W_LW_L}      & = \frac{\pi \alpha_2^2}{48 c_L^4 m_\psi^2} \left[ (2c_L^2 - 1)^2 - \frac{2 c_L s_R^2 s_L^2 (2c_L^2 - 1)}{c_R \sqrt{1-2s_L^2}}\left( 1 - \frac{m_{Z'}^2}{4m_\psi^2} \right)^{-1}  + \frac{c_L^2 s_L^4 s_R^4}{c_R^2(1-2s_L^2)} \left( 1 - \frac{m_{Z'}^2}{4m_\psi^2} \right)^{-2}\right],\nonumber \\
    \sigma v|_{\bar{f}f} & = \frac{\pi \alpha_2^2}{6 m_\psi^2} \left[ \frac{1}{c_L^4} \left( I_{3f}^2 - 2 I_{3f} s_L^2 Q_f + 2 s_L^4 Q_f^2 \right) \right. \nonumber                                                                                                                                                        \\
                         & \left. + \frac{2s_R^2}{c_R^2 c_L^2} \left( I_{3f}^2 s_R^2 - I_{3f} Q_f (s_R^2(s_L^2+1) + s_L^2) + 2 Q_f^2 s_L^2 s_R^2 \right) \left( 1 - \frac{m_{Z'}^2}{4m_\psi^2} \right)^{-1}\right. \nonumber                                                                                                \\
                         & \left.+ \frac{s_R^4}{c_R^4} \left( I_{3f}^2(1+s_R^4) - 2 I_{3f} Q_f (1+s_R^2) s_R^2 + 2 Q_f^2 s_R^4 \right) \left( 1 - \frac{m_{Z'}^2}{4m_\psi^2} \right)^{-2} \right],                                                                                                               \nonumber  \\
    \sigma v|_{Zh_L}       & = \frac{\pi \alpha_2^2}{48 c_L^2 m_\psi^2} \left[ \frac{1}{c_L^2} + \frac{2 s_L^2 s_R^2}{c_R c_L \sqrt{2 c_L^2 - 1}} \left( 1 - \frac{m_{Z'}^2}{4m_\psi^2} \right)^{-1} + \frac{s_R^4 s_L^4}{c_R^2 (2c_L^2 - 1)} \left( 1 - \frac{m_{Z'}^2}{4m_\psi^2} \right)^{-2} \right].
\end{align}

When the initial state $\psi_\ell^+ \psi_\ell^-$ total spin is 0, the annihilation cross sections are
\begin{align}
    \sigma v|_{\gamma \gamma} & = \frac{4 \pi  \alpha _2^2 s_L^4}{m_\psi^2}, ~~  \sigma v |_{ZZ}            = \frac{\pi  \alpha _2^2 \left(1-2 s_L^2\right)^4}{4 m_\psi^2 c_L^4}, ~~ \sigma v |_{Z \gamma}      = \frac{2 \pi  \alpha _2^2 s_L^2 \left(1-2 s_L^2\right)^2}{m_\psi^2 c_L^2},~~
    \sigma v |_{W_LW_L}           = \frac{\pi \alpha_2^2}{2 m_\psi^2},                                                  \nonumber                                                                                                                                                                 \\
    \sigma v |_{Z' \gamma}    & = \frac{2 \pi \alpha_2^2 s_R^4 s_L^2}{c_R^2 m_\psi^2} \left( 1 - \frac{m_{Z'}^2}{4m_\psi^2} \right),                                                                                                \nonumber                                                 \\
    \sigma v|_{Z Z'}          & = \frac{\pi \alpha_2^2 s_R^4 (1-2s_L^2)^2}{2 c_R^2 c_L^2 m_\psi^2} \left( 1 - \frac{m_{Z'}^2}{4m_\psi^2} \right)^3 \left( 1 - \frac{m_{Z'}^2}{4 m_\psi^2} + \frac{m_{Z'}^4}{32m_\psi^4} \right)^{-2}.
\end{align}

When the initial state $\psi_\ell^+ \psi_\ell^-$ total spin is 1, the annihilation cross sections are
\begin{align}
    \sigma v|_{W_LW_L}       & = \frac{\pi \alpha_2^2}{48 c_L^4 m_\psi^2} \left[ 1- \frac{2 c_L s_L^2s_R^2}{c_R \sqrt{2c_L^2 - 1}} \left( 1 - \frac{m_{Z'}^2}{4m_\psi^2} \right)^{-1} + \frac{c_L^2 s_L^4 s_R^4}{c_R^2 (2c_L^2 - 1)} \left( 1 - \frac{m_{Z'}^2}{4m_\psi^2} \right)^{-2} \right]\,, \nonumber                                                                                                                                                                                       \\
    \sigma v|_{Zh_L}       & = \frac{\pi \alpha_2^2}{48 c_R^2 c_L^2 m_\psi^2} \left[ \frac{c_R^2 (2c_L^2 - 1)^2}{c_L^2} - \frac{2 c_R s_R^2s_L^2 \sqrt{2c_L^2 - 1}}{c_L}\left( 1 - \frac{m_{Z'}^2}{4m_\psi^2} \right)^{-1} + \frac{s_L^4 s_R^4}{2c_L^2 - 1} \left( 1 - \frac{m_{Z'}^2}{4m_\psi^2} \right)^{-2} \right]\,,                                                                                                                                                              \nonumber \\
    \sigma v|_{\bar{f}f} & = \frac{\pi \alpha_2^2}{6 m_\psi^2} \left[ \frac{1}{c_L^4}\left( I_{3 f}^2 \left(1-2 s_L^2\right)^2 - 2 I_{3 f} Q_f \left(1-2s_L^2\right) s_L^2 \left(3-4 s_L^2\right) + 2 Q_f^2 s_L^4 (3-4s_L^2)^2\right) \right. \nonumber                                                                                                                                                                                                                        \\
                         & \left. - \frac{2 s_R^2}{c_R^2 c_L^2} \left( 1 - \frac{m_{Z'}^2}{4m_\psi^2} \right)^{-1} \left( 2 Q_f^2 s_R^2 s_L^2 \left(3-4 s_L^2\right)+I_{3 f}^2 s_R^2 \left(1 - 2 s_L^2\right) \right. \right. \nonumber                                                                                                                                                                                                                                              \\
                         & \left. \left. - I_{3 f} Q_f \left(s_R^2 \left(\left(2 c_L^2-1\right) s_L^2-2 s_L^4+1\right)+s_L^2 \left(3-4 s_L^2\right)\right) \right) \right. \nonumber                                                                                                                                                                                                                                                                                                   \\
                         & \left. + \frac{s_R^4}{c_R^4} \left( 2 Q_f^2 s_R^4-2 I_{3 f} Q_f \left(s_R^2+1\right) s_R^2+I_{3 f}^2 \left(s_R^4+1\right) \right) \left( 1 - \frac{m_{Z'}^2}{4m_\psi^2} \right)^{-2}\right]\,.
\end{align}

The initial states $\psi_\ell^0 \psi_\ell^0$ and $\psi_\ell^+ \psi_\ell^-$ can mix with each other.
The combined system has the following long-range potential,
\begin{align}
    V(r) = \begin{pmatrix}
               - \frac{\alpha_2}{4c_L^2 r}e^{-m_Z r} & -\frac{\alpha_2}{2r}e^{-m_{W_L} r}                                                            \\
               -\frac{\alpha_2}{2r}e^{-m_{W_L} r}        & 2 \delta m- \frac{\alpha}{r} - \frac{\alpha_2 (s_L^2 - \frac{1}{2})^2}{c_L^2 r}e^{-m_Z r}
           \end{pmatrix}\,,
\end{align}
in the $(\psi_\ell^0 \psi_\ell^0, \psi_\ell^+ \psi_\ell^-)$ basis.
The transition amplitude between these two initial states can be computed from the generalized optical theorem
\begin{align}
    \Gamma_{\psi_\ell^0 \psi_\ell^0 \leftrightarrow \psi_\ell^+ \psi_\ell^-} = \frac{1}{2}\sum_{\rm Final} \left((\sigma v|_{\psi_\ell^0 \psi_\ell^0 \to \rm Final}) (\sigma v|_{\psi_\ell^+ \psi_\ell^- \to \rm Final})\right)^{1/2}\,,
\end{align}
where the summation is over the common annihilation final states.

\subsubsection{$\bar{\psi}_l^0 \psi_\ell^-$ annihilation}
When the total spin of the initial state is 0, the annihilation cross sections are
\begin{align}
    \sigma v |_{W_LZ}        = \frac{\pi  \alpha _2^2 s_L^4}{m_\psi^2 c_L^2},~~
    \sigma v |_{W_L Z'}      = \frac{\pi \alpha _2^2 s_R^4}{c_R^2 m_\psi^2} \left( 1 - \frac{m_{Z'}^2}{4m_\psi^2} \right)\,,~~
    \sigma v |_{W_L \gamma}  = \frac{\pi  \alpha _2^2 s_L^2}{m_\psi^2}\,.
\end{align}

When the total spin of the initial state is 1, the annihilation cross sections are
\begin{align}
    \sigma v|_{W_LZ}         = \frac{\pi  \alpha _2^2}{24 m_\psi^2}\,,~~
    \sigma v|_{\bar{f}f'}  = \frac{\pi  \alpha _2^2}{6 m_\psi^2}\,,~~
    \sigma v|_{W_L h_L}       = \frac{\pi  \alpha _2^2}{24 m_\psi^2}\,.
\end{align}

The long-range effective potential is
\begin{align}
    V(r) =  \frac{\alpha_2}{4 c_L r}(1-2s_L^2) e^{-m_Z r}\,.
\end{align}

\subsection{Bi-doublet WIMP dark matter}

\subsubsection{$\psi^0 \bar{\psi}^0$ and $\psi^+ \psi^-$ mixed annihilation}

When the initial state $\psi^0 \bar{\psi}^0$ total spin is 0, the annihilation cross sections are
\begin{align}
    \sigma v |_{W_LW_L}   = \frac{\pi \alpha_2^2}{2 m_\psi^2}, ~~
    \sigma v |_{ZZ}   = \frac{\pi \alpha_2^2}{4 c_L^4 m_\psi^2},~~                                                                                                                                      
    \sigma v|_{Z Z'} & = \frac{\pi \alpha_2^2 c_R^2}{2 c_L^2 m_\psi^2} \left( 1 - \frac{m_{Z'}^2}{4m_\psi^2} \right)^3 \left( 1 - \frac{m_{Z'}^2}{4m_\psi^2} + \frac{m_{Z'}^4}{32 m_\psi^4} \right)^{-2}.
\end{align}

When the initial state $\psi^0 \bar{\psi}^0$ total spin is 1, the annihilation cross sections are
\begin{align}
    \sigma v |_{W_LW_L}      & = \frac{\pi \alpha_2^2}{48 c_L^4 m_\psi^2} \left[ (2c_L^2 - 1)^2 - \frac{2 c_L s_R^2 s_L^2 (2c_L^2 - 1)}{c_R \sqrt{1-2s_L^2}}\left( 1 - \frac{m_{Z'}^2}{4m_\psi^2} \right)^{-1}  + \frac{c_L^2 s_L^4 s_R^4}{c_R^2(1-2s_L^2)} \left( 1 - \frac{m_{Z'}^2}{4m_\psi^2} \right)^{-2}\right], \\
    \sigma v|_{\bar{f}f} & = \frac{\pi \alpha_2^2}{6 m_\psi^2} \left[ \frac{1}{c_L^4} \left( I_{3f}^2 - 2 I_{3f} s_L^2 Q_f + 2 s_L^4 Q_f^2 \right) \right. \nonumber                                                                                                                                               \\
                         & \left. + \frac{2}{c_L^2} \left( I_{3f}^2 s_R^2 - I_{3f} Q_f (s_R^2(s_L^2+1) + s_L^2) + 2 Q_f^2 s_L^2 s_R^2 \right) \left( 1 - \frac{m_{Z'}^2}{4m_\psi^2} \right)^{-1}\right. \nonumber                                                                                                  \\
                         & \left.+  \left( I_{3f}^2(1+s_R^4) - 2 I_{3f} Q_f (1+s_R^2) s_R^2 + 2 Q_f^2 s_R^4 \right) \left( 1 - \frac{m_{Z'}^2}{4m_\psi^2} \right)^{-2} \right],                                                                                                                                    \\
    \sigma v|_{Zh_L}       & = \frac{\pi \alpha_2^2}{48 c_L^2 m_\psi^2} \left[ \frac{1}{c_L^2} + \frac{2 s_L^2 c_R}{c_L \sqrt{2 c_L^2 - 1}} \left( 1 - \frac{m_{Z'}^2}{4m_\psi^2} \right)^{-1} + \frac{c_R^2 s_L^4}{ (2c_L^2 - 1)} \left( 1 - \frac{m_{Z'}^2}{4m_\psi^2} \right)^{-2} \right].
\end{align}

When the initial state $\psi^+ \psi^-$ total spin is 0, the annihilation cross sections are
\begin{align}
    \sigma v|_{\gamma \gamma} & = \frac{4 \pi  \alpha _2^2 s_L^4}{m_\psi^2},~~ \sigma v |_{ZZ}            = \frac{\pi  \alpha _2^2 \left(1-2 s_L^2\right)^4}{4 m_\psi^2 c_L^4},~~     \sigma v |_{Z \gamma}      = \frac{2 \pi  \alpha _2^2 s_L^2 \left(1-2 s_L^2\right)^2}{m_\psi^2 c_L^2},~~
    \sigma v |_{W_LW_L}            = \frac{\pi \alpha_2^2}{2 m_\psi^2},                                                                                                         \nonumber                                                                                                          \\
    \sigma v |_{Z' \gamma}    & = \frac{2 \pi \alpha_2^2 c_R^2 s_L^2}{m_\psi^2} \left( 1 - \frac{m_{Z'}^2}{4m_\psi^2} \right),                                                                                              ~~
    \sigma v|_{Z Z'}           = \frac{\pi \alpha_2^2 c_R^2 (1-2s_L^2)^2}{2 c_L^2 m_\psi^2} \left( 1 - \frac{m_{Z'}^2}{4m_\psi^2} \right)^3 \left( 1 - \frac{m_{Z'}^2}{4 m_\psi^2} + \frac{m_{Z'}^4}{32m_\psi^4} \right)^{-2}.
\end{align}

When the initial state $\psi^+ \psi^-$ total spin is 1, the annihilation cross sections are
\begin{align}
    \sigma v|_{W_LW_L}       & = \frac{\pi \alpha_2^2}{48 c_L^4 m_\psi^2} \left[ 1- \frac{2 c_L s_L^2 c_R}{\sqrt{2c_L^2 - 1}} \left( 1 - \frac{m_{Z'}^2}{4m_\psi^2} \right)^{-1} + \frac{c_L^2 s_L^4 c_R^2}{(2c_L^2 - 1)} \left( 1 - \frac{m_{Z'}^2}{4m_\psi^2} \right)^{-2} \right]\,, \nonumber                                                                                                                                                  \\
    \sigma v|_{Zh_L}       & = \frac{\pi \alpha_2^2}{48 c_L^4  m_\psi^2} \left[  (2c_L^2 - 1)^2 + 2 c_R c_L s_L^2\sqrt{2c_L^2 - 1} \left( 1 - \frac{m_{Z'}^2}{4m_\psi^2} \right)^{-1} + \frac{c_R^2 c_L^2 s_L^4}{2c_L^2 - 1} \left( 1 - \frac{m_{Z'}^2}{4m_\psi^2} \right)^{-2} \right]\,,                                                                                                                                             \nonumber \\
    \sigma v|_{\bar{f}f} & = \frac{\pi \alpha_2^2}{6 m_\psi^2} \left[ \frac{1}{c_L^4}\left( I_{3 f}^2 \left(1-2 s_L^2\right)^2 - 2 I_{3 f} Q_f \left(1-2s_L^2\right) s_L^2 \left(3-4 s_L^2\right) + 2 Q_f^2 s_L^4 (3-4 s_L^2)^2\right) \right. \nonumber                                                                                                                                                                        \\
                         & \left. - \frac{1}{c_L^2}\left( 1 - \frac{m_{Z'}^2}{4m_\psi^2} \right)^{-1} \left( 2 Q_f^2 s_R^2 s_L^2 \left(3-4 s_L^2\right)+I_{3 f}^2 s_R^2 \left(1 - 2 s_L^2\right) \right. \right. \nonumber                                                                                                                                                                                                           \\
                         & \left. \left. - I_{3 f} Q_f \left(s_R^2 \left(\left(2 c_L^2-1\right) s_L^2-2 s_L^4+1\right)+s_L^2 \left(2 c_L^2-2 s_L^2+1\right)\right) \right) \right. \nonumber                                                                                                                                                                                                                                                   \\
                         & \left. +  \left( 2 Q_f^2 s_R^4-2 I_{3 f} Q_f \left(s_R^2+1\right) s_R^2+I_{3 f}^2 \left(s_R^4+1\right) \right) \left( 1 - \frac{m_{Z'}^2}{4m_\psi^2} \right)^{-2}\right]\,.
\end{align}

The initial states $\psi^+ \psi^-$ and $\psi^0 \bar{\psi}^0$ can mix with each other. The combined system has the following long-range potential,
\begin{align}
    V(r) = \begin{pmatrix}
               2 \delta m - \frac{\alpha}{r} - \frac{\alpha_2}{r} \frac{(1-2s_L^2)^2}{4 c_L^2} e^{-m_Z r} & - \frac{\alpha_2}{2r} e^{-m_{W_L} r}      \\
               - \frac{\alpha_2}{2r} e^{-m_{W_L} r}                                                           & -\frac{\alpha_2}{4 c_L^2 r}e^{-m_Z r}
           \end{pmatrix},
\end{align}
in the $(\psi^+ \psi^-$, $\psi^0 \bar{\psi}^0)$ basis.

\subsubsection{$\psi^0 \psi^+$ annihilation}

When the total spin of the initial state is 0, the annihilation cross sections are
\begin{align}
    \sigma v|_{W_LZ}        = \frac{\pi \alpha_2^2 s_L^4}{c_L^2 m_\psi^2} \,,                                     ~~
    \sigma v|_{W_L Z'}      = \frac{\pi \alpha_2^2 c_R^2}{m_\psi^2} \left(1- \frac{m_{Z'}^2}{4 m_\psi^2}\right)\,, ~~
    \sigma v|_{W_L \gamma}  =\frac{\pi \alpha_2^2 s_L^2}{m_\psi^2}\,.
\end{align}

When the total spin of the initial state is 1, the annihilation cross sections are
\begin{align}
    \sigma v|_{W_LZ}         = \frac{\pi  \alpha _2^2}{24 m_\psi^2}\,, ~~
    \sigma v|_{\bar{f}f'}  = \frac{\pi  \alpha _2^2}{6 m_\psi^2}\,,  ~~
    \sigma v|_{W_L h_L}      = \frac{\pi  \alpha _2^2}{24 m_\psi^2}\,.
\end{align}

The long-range potential is
\begin{align}
    V(r) =  \frac{\alpha_2}{r} \frac{1-2s_L^2}{4 c_L^2} e^{-m_Z r}.
\end{align}

\subsubsection{$\psi^0 \psi^-$ annihilation}

When the total spin of the initial state is 0, the annihilation cross sections are
\begin{align}
    \sigma v|_{W_R \gamma} & = \frac{\pi  \alpha _2^2 s_L^2}{m_\psi^2} \left( 1 - \frac{m_{W_R}^2}{4m_\psi^2}\right),       ~
    \sigma v|_{W_R Z}       = \frac{\pi \alpha_2^2 c_L^2 }{c_L^2 m_\psi^2} \left( 1 - \frac{m_{W_R}^2}{4m_\psi^2} \right).
\end{align}

When the total spin of the initial state is 1, the annihilation cross sections are
\begin{align}
    \sigma v|_{\bar{f}f'} & = \frac{\pi \alpha_2^2}{6 m_\psi^2} \left(1 - \frac{m_{W_R}^2}{4m_\psi^2}\right)^{-2}\,,                                                                                                                 \nonumber            \\
    \sigma v |_{W_R Z}    & = \frac{\pi \alpha_2^2 s_L^4}{24 c_L^2 m_\psi^2} \left( \frac{m_Z}{m_{W_R}} \right)^2 \left(1 - \frac{m_{W_R}^2}{4m_\psi^2}\right)^{-1} \left( 1 + \frac{5 m_{W_R}^2}{2 m_\psi^2} + \frac{m_{W_R}^4}{16 m_\psi^4} \right) \,.
\end{align}

The long-range potential is
\begin{align}
    V(r) = - \frac{\alpha_2}{r} \frac{1-2s_L^2}{4 c_L^2} e^{-m_Z r}.
\end{align}

\subsubsection{$\psi^0 \psi^0$ annihilation}
The initial state should have total spin 0. The only possible final state is
\begin{align}
    \sigma v |_{W_L W_R} = \frac{2 \pi \alpha_2^2}{m_\psi^2} \left( 1 - \frac{m_{W_R}^2}{4m_\psi^2} \right) \left( 1 - \frac{m_{W_R}^2}{4 m_\psi^2} + \frac{m_{W_R}^4}{32 m_\psi^4} \right)^{-2}\,.
\end{align}

The long-range potential is
\begin{align}
    V(r) = \frac{\alpha_2}{4 c_L^2 r}e^{-m_Z r}\,.
\end{align}

\subsubsection{$\psi^+ \psi^+$ annihilation}
The initial state should have total spin 0. The only possible final state is
\begin{align}
    \sigma v |_{W_L W_R} = \frac{2 \pi \alpha_2^2}{m_\psi^2} \left( 1 - \frac{m_{W_R}^2}{4m_\psi^2} \right) \left( 1 - \frac{m_{W_R}^2}{4 m_\psi^2} + \frac{m_{W_R}^4}{32 m_\psi^4} \right)^{-2}\,.
\end{align}

The long-range potential is
\begin{align}
    V(r) = \frac{\alpha}{r} +  \frac{\alpha_2}{r} \frac{(1-2s_L^2)^2}{4 c_L^2} e^{-m_Z r}\,.
\end{align}

\section{Decay rate of $Z'$ and $W_R$}

In this appendix, we discuss the heavy gauge boson decays.
For conciseness and readability, we ignore the final state masses when presenting results here, but we include the masses for $m_t$, $m_{W_L}$, $m_Z$, and $m_{h_L}$ in our computations.
The masses of new fermions are treated in the same way as described in Sec.~\ref{sec:annihilation}.

$W_R$ can decay into pairs of SM fermions or $\psi_r^- \psi_r^0$.
The total decay rate is
\begin{align}
    \Gamma_{W_R} = 12 \Gamma_{W_R\to \rm SM} + \Gamma_{W_R \to \psi},
\end{align}
with
\begin{align}
    \Gamma_{W_R\to \rm SM}             & = \frac{\alpha m_{W_R}}{12 s_L^2}      \,,                                                                                                           \\
    \Gamma_{W_R \to \psi_r^- \psi_r^0} & = \Gamma_{W_R \to \psi^- \psi^0} = \frac{\alpha_2}{6} m_{W_R} \sqrt{1-\frac{4 m_\psi^2}{m_{W_R}^2}} \left( 1 - \frac{m_\psi^2}{m_{W_R}^2} \right)\,,
\end{align}
where $W_R$ can decay into $\psi_r^- \psi_r^0$ in the doublet-pair fermion while into $\psi^- \psi^0$ for the bi-doublet fermion.

On the other hand, $Z'$ can decay into SM fermion pairs, $Zh_L$, and extra fermions.
In addition, $Z'$ can decay into $W_L^+ W_L^-$ via mixing with $Z$.
The total decay rate is thus
\begin{align}
    \Gamma_{Z'} = \sum_f \Gamma_{Z' \to \rm SM} + \Gamma_{Z' \to \psi},
\end{align}
with
\begin{align}
    \Gamma_{Z'\to \bar{f}f}                 & = \frac{\alpha_2 m_{Z'}}{6 c_R^2} \left( I_{3f}^2 (1+s_R^4) - 2 I_{3f} Q_f (1+s_R^2)s_R^2 + 2 Q_f^2 s_R^4 \right),                                         \\
    \Gamma_{Z'\to Zh_L}                       & = \frac{\alpha_2 c_L^2 s_R^4 \left(c_R^2+2 s_R^2\right) m_{Z_R}}{24  c_R^2},                                                                               \\
    \Gamma_{Z' \to {W_L}{W_L}}                      & = \frac{\alpha_2 s_L^4 m_{Z'}}{48  c_L^2 (1-2s_L^2)}                                                                                                      ,\\
    \Gamma_{Z' \to \psi_r^+ \psi_r^-}       & = \frac{\alpha_2 (1-2s_R^2)^2 m_{Z'}}{12 c_R^2} \sqrt{1-\frac{4m_\psi^2}{m_{Z'}^2}} \left( 1 - \frac{m_\psi^2}{m_{Z'}^2} \right)                          ,\\
    \Gamma_{Z' \to \bar{\psi}_r^0 \psi_r^0} & = \frac{\alpha_2 m_{Z'}}{12 c_R^2} \sqrt{1-\frac{4m_\psi^2}{m_{Z'}^2}} \left( 1 - \frac{m_\psi^2}{m_{Z'}^2} \right)                                       ,\\
    \Gamma_{Z' \to \psi_\ell \psi_\ell}     & = \frac{\alpha_2 s_R^4 m_{Z'}}{12 c_R^2} \sqrt{1-\frac{4m_\psi^2}{m_{Z'}^2}} \left( 1 - \frac{m_\psi^2}{m_{Z'}^2} \right)                                 ,\\
    \Gamma_{Z' \to \bar{\Psi}^0 \Psi^0}     & = \Gamma_{Z' \to \Psi^+ \Psi^-} = \frac{\alpha_2^2 c_R^2 m_{Z'}}{12 \pi} \sqrt{1-\frac{4m_\psi^2}{m_{Z'}^2}} \left( 1 - \frac{m_\psi^2}{m_{Z'}^2} \right),
\end{align}
where $\Gamma_{Z' \to \psi_r^+ \psi_r^-}$, $\Gamma_{Z' \to \bar{\psi}_r^0 \psi_r^0}$ and $\Gamma_{Z' \to \psi_\ell \psi_\ell}$ (including two flavors) are relevant to the doublet pair WIMP model, and $\Gamma_{Z' \to \bar{\Psi}^0 \Psi^0}$ and $ \Gamma_{Z' \to \Psi^+ \Psi^-}$ are relevant to the bi-doublet WIMP model.

\small{\bibliography{DM_parity_minimal.bib}}

\end{document}